\def\year{2022}\relax
\documentclass[letterpaper]{article} 
\usepackage{aaai22}  
\usepackage{times}  
\usepackage{helvet}  
\usepackage{courier}  
\usepackage[hyphens]{url}  
\usepackage{graphicx} 
\usepackage{natbib}  
\usepackage{caption} 
\DeclareCaptionStyle{ruled}{labelfont=normalfont,labelsep=colon,strut=off} 
\usepackage{algorithm}
\usepackage{algorithmic}

\usepackage{newfloat}
\usepackage{listings}
\floatstyle{ruled}
\newfloat{listing}{tb}{lst}{}
\floatname{listing}{Listing}

\usepackage{multirow}
\usepackage{hyperref} 

\title{Versatile Learned Video Compression}
\author {
    Runsen Feng\textsuperscript{\rm 1},
    Zongyu Guo\textsuperscript{\rm 1},
    Zhizheng Zhang\textsuperscript{\rm 2},
    Weiping Li\textsuperscript{\rm 1},
    Zhibo Chen\textsuperscript{\rm 1}\thanks{Corresponding author.}
}
\affiliations {
    \textsuperscript{\rm 1} University of Science and Technology of China\\
    \textsuperscript{\rm 2} Microsoft Research Asia\\
    \{fengruns, guozy\}@mail.ustc.edu.cn, zhizzhang@microsoft.com, \{wpli, chenzhibo\}@ustc.edu.cn
}

\usepackage{amsmath}
\usepackage{subcaption}
\usepackage{cleveref}

\begin{document}

\maketitle

\begin{abstract}
Learned video compression methods have demonstrated great promise in catching up with traditional video codecs in their rate-distortion (R-D) performance. However, existing learned video compression schemes are limited by the binding of the prediction mode and the fixed network framework. They are unable to support various inter prediction modes and thus inapplicable for various scenarios.
In this paper, to break this limitation, we propose a versatile learned video compression (VLVC) framework that uses one model to support all possible prediction modes.
Specifically, to realize versatile compression, we first build a motion compensation module that applies multiple 3D motion vector fields (\textit{i.e.}, voxel flows) for weighted trilinear warping in spatial-temporal space. The voxel flows convey the information of temporal reference position that helps to decouple inter prediction modes away from framework designing.
Secondly, in case of multiple-reference-frame prediction, we apply a flow prediction module to predict accurate motion trajectories with unified polynomial functions. We show that the flow prediction module can largely reduce the transmission cost of voxel flows.
Experimental results demonstrate that our proposed VLVC not only supports versatile compression in various settings, but also is the first end-to-end learned video compression method that outperforms the latest VVC/H.266 standard reference software in terms of MS-SSIM.
\end{abstract}

\section{Introduction}
Video occupies more than 80\% of network traffic and the amount of video data is increasing rapidly \cite{cisco2018cisco}. Thus, the storage and transmission of video become more challenging. 
A series of hybrid video coding standards have been proposed, such as AVC/H.264~\cite{wiegand2003overview}, HEVC/H.265~\cite{sullivan2012overview} and the latest video coding standard VVC/H.266~\cite{vvc}. These traditional standards are manually designed and the development of the compression framework is gradually saturated. 
Recently, the performance of video compression is mainly improved by designing more complex prediction modes, leading to increased coding complexity.

Deep neural networks are currently promoting the development of data compression.
Despite the remarkable progress on the field of learned image compression \cite{balle2016end,balle2018variational,minnen2018joint,cheng2020learned,agustsson2020universally,guo2021soft}, the area of learned video compression is still in early stages. 
Existing methods for learned video compression can be grouped into three categories, including frame interpolation-based methods \cite{wu2018video, djelouah2019neural}, 3D autoencoder-based methods \cite{habibian2019video, liu2020conditional}, and predictive coding methods with optical flow such as \cite{lu2019dvc, Agustsson_2020_CVPR}. 
So far, among them, video compression with optical flow presents the best performance \cite{rippel2021elf}, where the optical flow represents a pixel-wise motion vector (MV) field utilized for inter frame prediction. In this paper, we also focus on this predictive coding architecture.
Previous works with optical flow are proposed to support specific 
prediction mode, including unidirectional or bidirectional, single or multiple frame prediction. 
They are too cumbersome to support versatile compression in various settings since they bind the inter prediction mode with the fixed network framework.

It is important to design a more flexible model to handle all possible settings like traditional codecs. For example, the lowdelay configurations (coding with unidirectional reference) are effective for the scenarios such as live streaming which requires low latency coding. However, these configurations are less applicable for the randomaccess scenarios like playback which requires the fast decoding of arbitrary target frames. Therefore, the randomaccess configurations (coding with bidirectional reference) are gravely needed for these scenarios.
In this paper, we propose a versatile learned video compression (VLVC) framework that achieves coding flexibility as well as compression performance. A voxel flow based motion compensation module is adopted for higher flexibility, which is then extended into multiple voxel flows to perform weighted trilinear warping. In addition, in case of multiple-reference-frame prediction, a polynomial motion trajectories based flow prediction module is designed for better compression performance. Our motivations are as follows.

\begin{figure*}[t]
\centering
\subfloat[\label{fig1a}]{
    \includegraphics[width=0.129\linewidth]{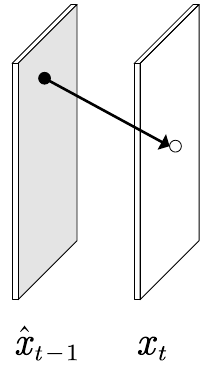}
}\hfill
\subfloat[\label{fig1b}]{
    \includegraphics[width=0.20\linewidth]{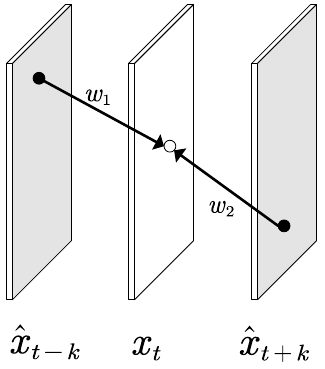}
}\hfill
\subfloat[\label{fig1c}]{
    \includegraphics[width=0.20\linewidth]{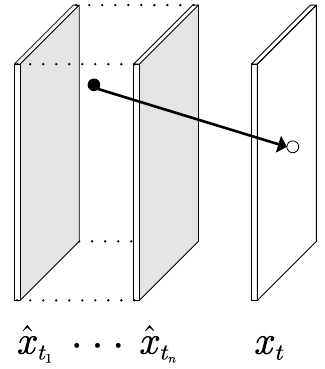}
}\hfill
\subfloat[\label{fig1d}]{
    \includegraphics[width=0.20\linewidth]{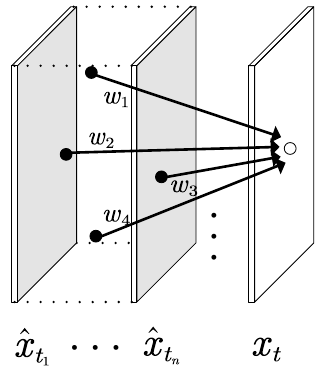}
}
\caption{Different motion compensation (inter frame prediction) methods. (a) Unidirectional prediction with 2D optical flow \cite{lu2019dvc}. (b) Bidirectional prediction with two optical flows and weight coefficients \cite{djelouah2019neural}. (c) Prediction with a single voxel flow \cite{liu2017video}, freely sampling the reference frames in spatial-temporal space. (d) Prediction with multiple voxel flows via weighted trilinear warping.}
\label{fig1}
\end{figure*}

\textbf{Motion compensation with multiple voxel flows.}
Previous works such as \cite{lu2019dvc} apply 2D optical flow for low-delay prediction using a single reference frame (unidirectional prediction, see Fig.~\ref{fig1a}). 
For the practical random access scenario, bidirectional reference frames are available for more accurate frame interpolation \cite{djelouah2019neural} (Fig.~\ref{fig1b}).
However, the reference positions in these works are determined by pre-defined prediction modes. They cannot adapt to various inter prediction modes where reference positions are various. 
In this paper, we apply 3D voxel flows to describe not only the spatial MVs, but also the information of temporal reference positions (Fig.~\ref{fig1c} \& Fig.~\ref{fig1d}). 
We perform voxel flow based motion compensation via trilinear warping, which is applicable to single or multiple, unidirectional or bidirectional reference frames.
Unlike \cite{Agustsson_2020_CVPR} that adopts scale space flow with trilinear warping, we apply voxel flows for inter prediction in spatial-temporal space, which naturally renders our model more robust to different coding scenarios. 
Furthermore, beyond using a single MV in every position of the current frame, we propose to use multiple voxel flows to describe multiple possible reference relationships (Fig.~\ref{fig1d}). 
Then the target pixel is synthesized as the weighted fusing of warping results. 
We show that compared to the single voxel flow based warping, the proposed weighted warping with multiple voxel flows is more accurate, yielding less residual and more efficient compression.

\textbf{Flow prediction with polynomial motion trajectories.}
Exploiting multiple reference frames usually achieves better compression performance since more reference information is provided.
A versatile learned video compression model should cover this multi-reference case. While previous work \cite{lin2020m} designs a complex flow prediction network to reduce the redundancies of 2D MV fields, the number and structure of reference frames are inherent and fixed within the framework.
In this paper, we design a more intelligent method for flow prediction, \textit{i.e.}, modeling the prediction modes with polynomial coefficients. We formulate different motion trajectories in a time interval by a unified polynomial function. 
The polynomial coefficients are solved by establishing a multivariate equation (see Section \ref{sec:GMP}).
Since this polynomial function models the accurate motion trajectories, it serves as a basic discipline that constrains the predicted motion to be reasonable. We show the transmission cost of voxel flows is reduced obviously with the help of additional motion trajectory information.

Thanks to the above two technical contributions, our proposed VLVC is not only applicable for various practical compression scenarios with different inter prediction modes, but also delivers impressive R-D performance on standard test sequences. Extensive experimental results demonstrate that VLVC
is the first learning-based method to 
outperform the Versatile Video Coding (VVC) standard in terms of MS-SSIM in both low delay and random access configurations. Comprehensive ablation studies and discussions are provided to verify the effectiveness of our method.

The remainder of this paper is organized as follows. In the next section, we briefly overview some related works. In Section~\ref{sec:VLVC}, we introduce the proposed versatile learned video compression framework and provide detailed descriptions of the voxel flow based warping and polynomial motion modeling. The experimental results and analysis will be provided in Section~\ref{sec:experiments}. Finally, we conclude this paper in Section~\ref{sec:conclusion}.

\section{Related Work}\label{sec:related_work}
\paragraph{Learned Image Compression} 
Recent advances in learned image compression~\cite{balle2016end,balle2018variational,minnen2018joint}, have shown the great success of nonlinear transform coding. Many existing methods are built upon hyperprior-based coding framework~\cite{balle2018variational}, which are improved with more efficient entropy models~\cite{minnen2018joint,cheng2020learned}, variable-rate compression~\cite{cui2020g} and more effective quantization~\cite{agustsson2020universally,guo2021soft}. While the widely used autoregressive entropy models provide significant performance gain in image coding, the high decoding complexity is not suitable for practical video compression. We thus only employ the hyperprior model~\cite{balle2018variational} as the entropy model in our video coding framework.

\paragraph{Learned Video Compression}
As mentioned before, existing approaches can be divided into three categories. 3D autoencoder-based methods \cite{habibian2019video, liu2020conditional}, as the video extensions of nonlinear transform coding \cite{balle2020nonlinear}, aim to transform video into a quantized spatial-temporal representation. However, they are currently much inferior in performance, compared with the other two categories. 
The other two categories follow a similar coding pipeline: first perform inter-frame prediction using either backward warping operation or frame interpolation networks, and then compress the corresponding residual information using autoencoder-based networks.
For example, \cite{chen2019learning} propose a spatial-recurrent compression framework in block level. \cite{lu2019dvc} propose a fully end-to-end trainable framework, where all key components in the classical video codec are implemented with neural networks. \cite{djelouah2019neural} perform interpolation by the decoded optical flow and blending coefficients. They reuse the same autoencoder of I-frame compression and directly quantize the corresponding latent space residual. \cite{2020HLVC} propose a video compression framework with three hierarchical quality layers and recurrent enhancement. In \cite{lin2020m}, multiple frames motion prediction are introduced into the P-frame coding. \cite{Agustsson_2020_CVPR} replace the bilinear warping operation with scale-space flow which learns to adaptively blur the reference content for better warping results. However, most existing methods are designed for particular prediction modes, resulting in inflexibility for different scenarios. The recent work of \cite{ladune2021conditional} applies a weight map to adaptively determine the P frame or B frame prediction. But it is limited to only two reference frames and cannot deal with more complex reference structures.

\paragraph{Video Interpolation}
The task of video interpolation is closely related to video compression. One pioneering work \cite{liu2017video} proposes to use deep voxel flow to synthesize new video frames. Some works of video interpolation \cite{niklaus2017video, reda2018sdc, bao2019memc} directly generate the spatially-adaptive convolutional kernels. Most recently, \cite{lee2020adacof, TMMshi2021video} proposed to relax the kernel shape, to select multiple sampling points freely in space or space-time. In this paper, our proposed multiple voxel flows based warping is motivated by the accurate interpolation results in \cite{liu2017video, TMMshi2021video}. Some recent works~\cite{B-EPIC, TMMliu2019deep} for video compression directly employ deep video interpolation to generate a better reference frame for inter prediction. However, their video interpolation networks are designed for fixed reference structures and are inflexible for various prediction modes.

\paragraph{Optical Flow Estimation}
Compared to the task-oriented motion descriptors used in video interpolation and video compression, optical flow is more fundamental and robust visual information, which is suitable for various multimedia tasks like action recognition~\cite{TMMshi2017sequential}, video compression~\cite{TMM2014Fast, lu2019dvc}, video super-resolution~\cite{caballero2017real} and so on. In this paper, we choose optical flow as the base motion descriptor to build a generic motion model which is not sensitive to quantization noise. Due to the great success of deep learning-based optical flow estimation~\cite{TMM2019Refined, sun2018pwc, TMM2018Recurrent, TMM2018selfattn}, we here employ PWC-net~\cite{sun2018pwc} as the optical flow estimator in our generalized flow prediction module.

\begin{figure*}[htb]
\centering
\includegraphics[width=\linewidth]{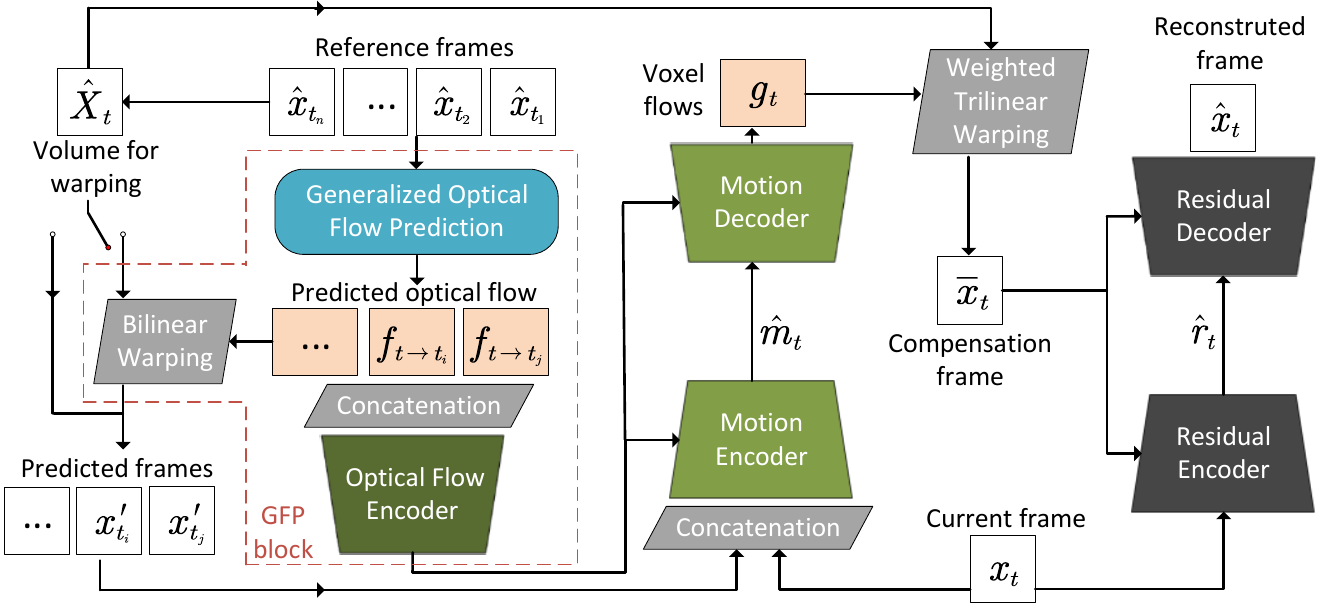}
\caption{The overall architecture of our VLVC framework: 1) the generalized flow prediction (GFP) block included in the red dashed box models polynomial motion trajectories based on the transmitted reference frames, and it then predicts optical flow from several reference frames to current frame; 2) conditioned on the predicted optical flow, the voxel flows is jointly estimated and compressed into quantized latents $\boldsymbol{\hat{m}}_t$; 3) when the GFP block is turned off, the predicted optical flow is set to zero and the predicted frames are identical to the reference volume. The detailed structures can be found in Section~\ref{sec:experiments}.}
\label{fig:overview}
\end{figure*}

\section{Versatile Learned Video Compression}\label{sec:VLVC}
\textbf{Notations.} To compress videos, the original video sequence is first divided into groups of pictures (GOP). Let $\boldsymbol{x} = \{\boldsymbol{x}_1,\boldsymbol{x}_2, ...,  \boldsymbol{x}_T\}$ refer to the frames in one GOP unit where the GOP size is $T$. 
To take advantage of previous decoded frames, our model predicts the current frame $\boldsymbol{x}_{t}$ with $\boldsymbol{n}$ reference frame(s), \textit{i.e.}, the lossy reconstruction results compared to the original frames.
Here, we denote the reference frames as $\{\boldsymbol{\hat{x}}_{t_1},\boldsymbol{\hat{x}}_{t_2}, ...,  \boldsymbol{\hat{x}}_{t_n}  \}$, where $\{t_1, t_2, ..., t_n\}$ is the index of temporal reference position.
If multiple frames are taken as the references (\textit{i.e.}, $n>1$), these previously reconstructed frames can be divided into two groups: one is used only for flow prediction to reduce the transmission cost of motion information, and the other is used for both flow prediction and motion compensation (warping).
In other words, the reference frames which are directly taken for warping are from a sub-set of $\{\boldsymbol{\hat{x}}_{t_1},\boldsymbol{\hat{x}}_{t_2}, ...,  \boldsymbol{\hat{x}}_{t_n}\}$, which could be stacked into a volume denoted by $\boldsymbol{\hat{X}}_t$. If only one previous reference frame is available, the volume for warping is $\boldsymbol{\hat{X}}_t = \{\boldsymbol{\hat{x}}_{t-1} \}$.

An overview of our video compression framework is shown in Fig. \ref{fig:overview}. In short, our model contains a motion encoder and decoder, a residual encoder and decoder, and a flow prediction module.
Both the motion encoder/decoder and the residual encoder/decoder are similar to autoencoder-based image compression network \cite{balle2016end,balle2018variational}. Note that one previous work \cite{Agustsson_2020_CVPR} demonstrates that a pre-trained flow extractor is unnecessary for motion encoder, which is followed by us. Therefore, if we do not consider our proposed flow prediction module, our video compression model is similar as \cite{Agustsson_2020_CVPR}, except for the input of motion encoder and the output of motion decoder. In our framework, the motion encoder is fed with the current frame $\boldsymbol{x}_t$ concatenated with the predicted frames (represented as $\boldsymbol{x}_t'$ in Fig. \ref{fig:overview}). Here, each predicted frame $\boldsymbol{x}_t'$ is an estimation for the current frame $\boldsymbol{x}_t$. All these predicted frames reveal how much information the decoder already knows about the current frame. 
On the decoder side, the motion decoder will generate multiple voxel flows for more effective motion compensation. The details of such motion compensation mechanism are explained in Section \ref{sec:MMCVF}.

The generalized flow prediction (GFP) block (included in the red dashed box in Fig. \ref{fig:overview}) is turned off when only a single reference frame is available. And the GFP block can be turned on for multiple-reference prediction, which is usually applied in the scenario that allows higher computational complexity and higher latency. As shown in Fig. \ref{fig:overview}, the GFP block not only generates the predicted frames, which are taken as a part of the input of motion encoder, but also provides extra auxiliary information for motion encoding and decoding. The auxiliary information here is modeled as motion trajectories, helping the motion encoder/decoder compress motion information effectively. As a result, it is able to reduce the transmission cost of the quantized motion information $\boldsymbol{\hat{m}}_t$. The specific introduction of our proposed flow prediction module can be found in Section \ref{sec:GMP}.

\subsection{Prediction with multiple voxel flows}\label{sec:MMCVF}
Voxel flow \cite{liu2017video} is a per-pixel 3-D motion vector that describes relationships in spatial-temporal domain. 
Compared to 2D optical flow, voxel flow can inherently allow the codec to be aware of the sampling positions in the temporal dimension for various prediction modes.
Given arbitrary reference frames, the model is expected to select the optimal reference position for better reconstructing the current frame to be compressed. 
Such a 3-D motion descriptor helps to build a prediction-mode-agnostic video coding framework, \textit{i.e.}, a versatile learned video codec.

In addition, a single flow field is hard to represent complex motion (e.g. blurry motion), which may result in inaccurate prediction and high coding cost of residuals.
On the other hand, a local region can be predicted with multiple reference sources.
Thereby, in this work, we propose to use multiple voxel flows to perform weighted trilinear warping by sampling in $\boldsymbol{X}_t$ for multiple times. We remind our readers that $\boldsymbol{X}_t$ is a volume consisting of some reference frames. 
Assume the dimension of $\boldsymbol{X}_t$ is $ D  \times H \times W$, where $D$ is the number of reference frames used for motion compensation. the motion decoder will generate multiple voxel flows by outputting a $ (4M) \times H \times W$ tensor. Here, $M$ refers to the flow number. Therefore, every voxel flow is a 4-channel field describing the 3-channel voxel flow $\boldsymbol{g}^i = (\boldsymbol{g}^i_{x}, \boldsymbol{g}^i_{y}, \boldsymbol{g}^i_{z})$ with a corresponding weight channel $\boldsymbol{g}^i_{w}$. Here, $i$ ($1 \leq i \leq M$) is the index of voxel flow.
To synthesize the target pixels in current frame, the weights $\boldsymbol{g}^i_{w}$ are normalized by a softmax function across $M$ voxel flows. We finally obtain the target pixel $\boldsymbol{\bar{x}}[x, y]$ in spatial location $[x, y]$ by calculating the weighted sum of sampling results, formulated as:

\begin{equation}\label{eq:7}
\begin{aligned}
\boldsymbol{\bar{x}}[x, y] = \sum_{i=1}^M g^i_w(x, y) \boldsymbol{X}_t[ 
 & x + g^i_x(x, y), \\
& y + g^i_y(x, y), g^i_z(x, y)].
\end{aligned}
\end{equation}

We experimentally find that compared with a single voxel flow, the extra transmission cost of multiple voxel flows is negligible. The model is able to learn appropriate motion information under the rate-distortion optimization. In other words, the model is optimized to avoid the transmission of unnecessary flows. Meanwhile, the bits consumed by residuals decrease obviously with the help of our proposed prediction method using multiple voxel flows.

\subsection{Generalized optical flow prediction}\label{sec:GMP}
In the VLVC framework, a generalized flow prediction module is proposed to reduce the transmission cost of temporal-consecutive voxel flows. This module can be turned on in case of prediction with multiple reference frames. While one previous work involving multiple-reference prediction \cite{lin2020m} fixes the prediction mode into a complex flow prediction module, it cannot deal with different prediction settings, no matter unidirectional prediction with various reference structures or bidirectional prediction. Here, we explore to explicitly model the temporal motion trajectory with a polynomial function that can adapt to all prediction modes.
By solving the polynomial coefficients and then reversing the flow (introduced later), our model can estimate the predicted optical flow from a selected reference frame to the current frame. Such predicted optical flow is available on both the encoder and decoder sides. On the one hand, the predicted optical flow is used to generate the predicted frame, which is a part of the input of motion encoder/decoder. On the other hand, the predicted optical flow will be integrated into motion encoding/decoding as additional auxiliary information, to facilitate the compression of voxel flows.

Now we introduce how to generate the predicted optical flow. 
First of all, we should note that there are two kinds of optical flows describing the motion between the reference frame $\boldsymbol{\hat{x}}_{t_j}$ and the target frame $\boldsymbol{x}_t$: forward flow $f_{t_j \rightarrow t}$ and backward flow $f_{t \rightarrow t_j}$ \cite{xu2019quadratic,niklaus2020softmax}. Our target is to predict the backward flow, which can be utilized to sample $\boldsymbol{\hat{x}}_{t_j}$ to generate corresponding pixels in the target frame $\boldsymbol{x}_t$ via backward warping. However, since the target frame is unavailable on the decoder side, direct prediction of backward flow is difficult. Therefore, we choose to first model the motion trajectory of reference pixels to predict forward flow and then reverse the forward flow to predict the backward flow\cite{xu2019quadratic}.

\begin{figure}[tb]\label{fig3}
\centering
\includegraphics[width=\linewidth]{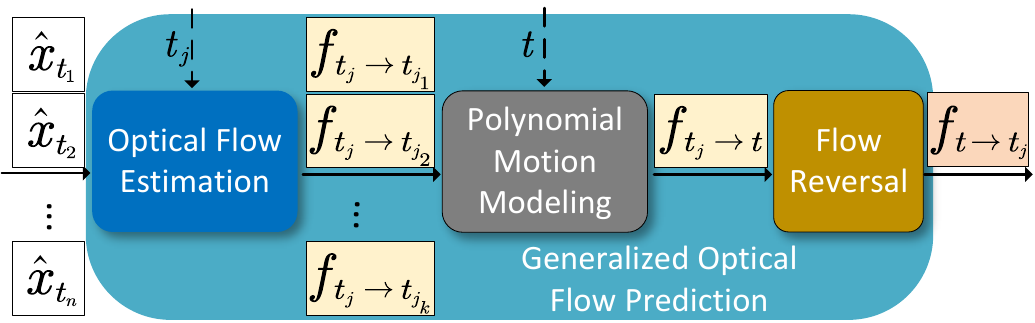}
\caption{Generalized optical flow prediction module. To predict the backward flow from $t_j$ to $t$ with arbitrary reference frames, the module uses several estimated optical flow to solve the coefficients of polynomial motion functions, and then it predicts and reverses the forward flow to get the final prediction.}
\end{figure}

\paragraph{Polynomial motion modeling with forward flows} 
We first select a temporal reference stamp $t_j$ as the origin of the reference coordinate system. For each pixel at $t_j$, we model its forward motion $f_{t_j \rightarrow t}$ by the $k$-order ($k<n$) polynomial functions within this reference coordinate system:

\begin{equation}\label{eq:1}
\centering
\begin{aligned}
f_{t_j \rightarrow t}
= \sum_{l=1}^k a_l \times (t - t_j)^l,
\end{aligned}
\end{equation}
where $a_1, ..., a_k$ are polynomial coefficients for modeling the motion of a pixel. The reference origin $t_j$ is selected among the time stamps of reference frames $\{\boldsymbol{\hat{x}}_{t_1},\boldsymbol{\hat{x}}_{t_2}, ...,  \boldsymbol{\hat{x}}_{t_n}  \}$. And we can solve the polynomial coefficients in Eq. \ref{eq:1} by setting $t$ equal to the top-$k$ nearest time stamp $\{t_{j_i}\}_{i=1}^k$ around $t_j$, which are also in the set of reference time stamps. Then we can obtain the equation:

\begin{small}
\begin{equation}\label{eq:2}
\centering
\begin{aligned}
\mathbf{A}
 =
 \begin{bmatrix}
   (t_{j_1} - t_j) & (t_{j_1} - t_j)^2 & ... &(t_{j_1} - t_j)^k \\
   (t_{j_2} - t_j) & (t_{j_2} - t_j)^2 & ... &(t_{j_2} - t_j)^k \\
   ... & ... & ... & ...\\
   (t_{j_{k}} - t_j) & (t_{j_k} - t_j)^2 & ... & (t_{j_{k}} - t_j)^k \\
 \end{bmatrix}^{-1} 
 \mathbf{F},
\end{aligned}
\end{equation}
\end{small}
where $\mathbf{A} = [a_1, ..., a_k]^{\mathrm{T}}$ is the coefficients matrix, and $\mathbf{F} = [f_{t_j \rightarrow t_{j_1}}, f_{t_j \rightarrow t_{j_2}}, ..., f_{t_j \rightarrow t_{j_{k}}}]^{\mathrm{T}}$ can be obtained using off-the-shelf flow estimation network, such as a pre-trained PWC-Net \cite{sun2018pwc} in our work. Then we can derive the polynomial coefficients and apply them to  Eq.~(\ref{eq:1}) to predict the forward flow from $t_j$ to any time stamp $t$. 

\begin{figure*}[htb]
\centering 
\subfloat{
    \includegraphics[width=0.315\linewidth]{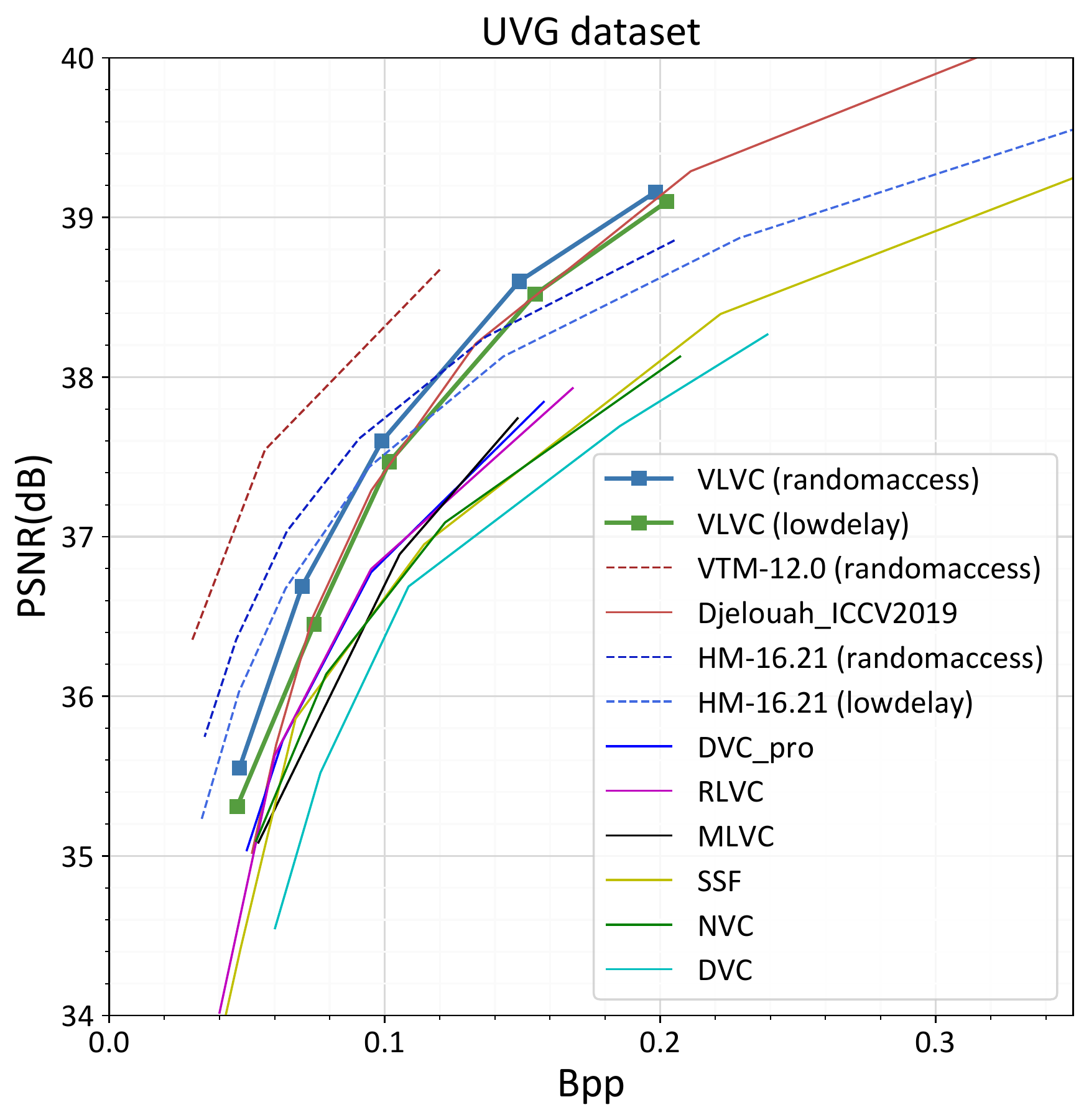}
}\hfill
\subfloat{
    \includegraphics[width=0.32\linewidth]{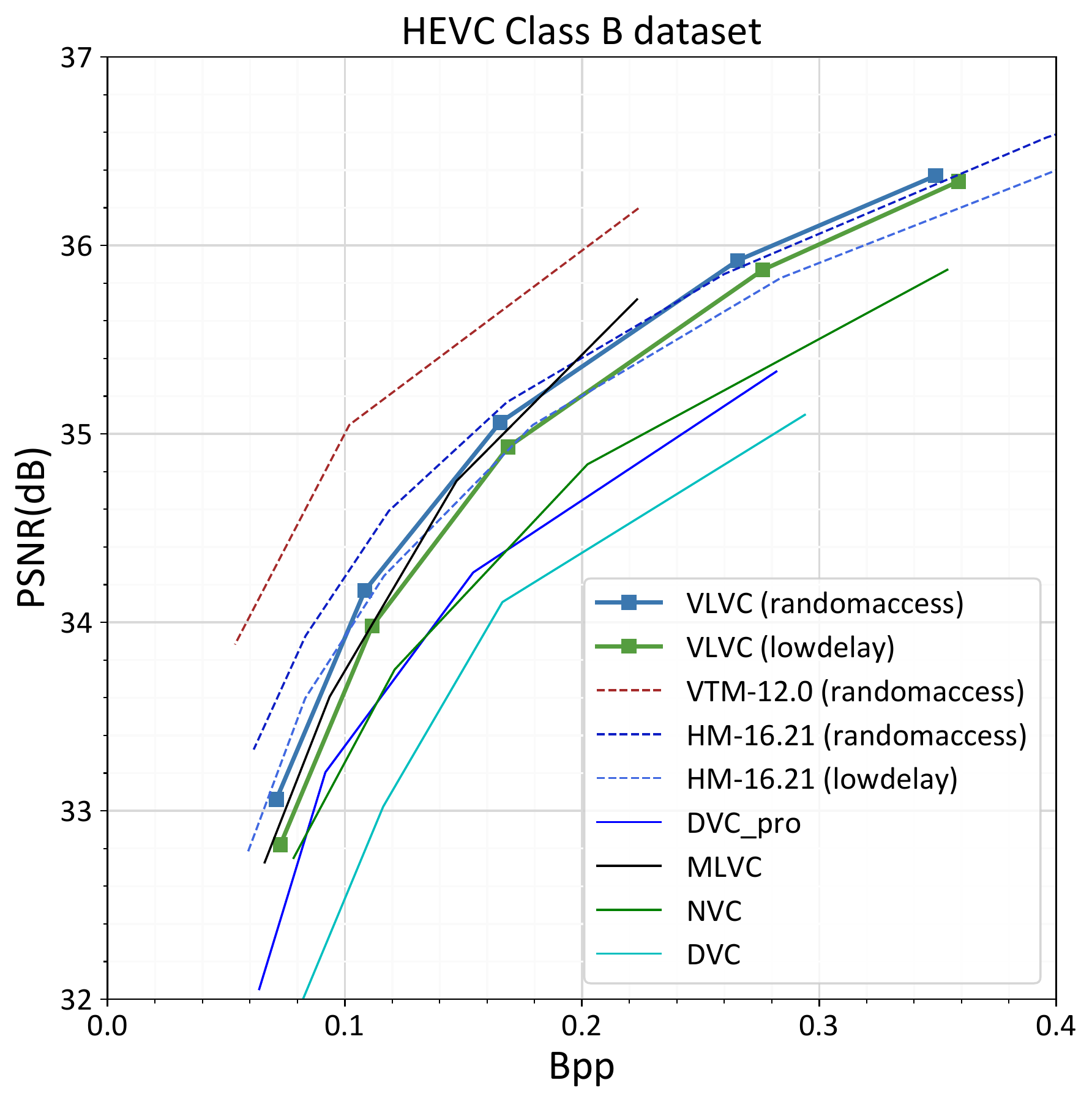}
}\hfill
\subfloat{
    \includegraphics[width=0.323\linewidth]{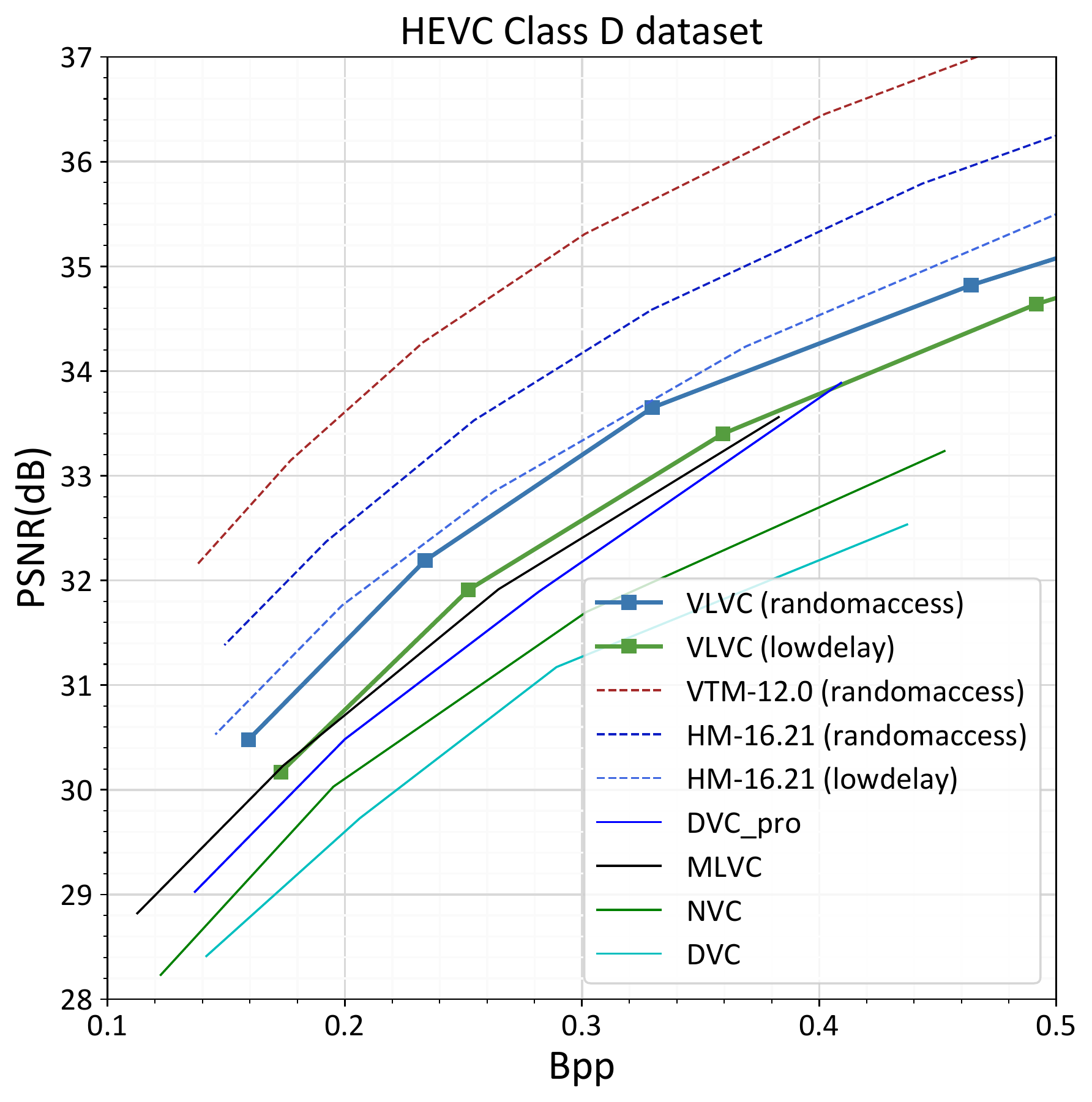}
}\\
\subfloat{
    \includegraphics[width=0.315\linewidth]{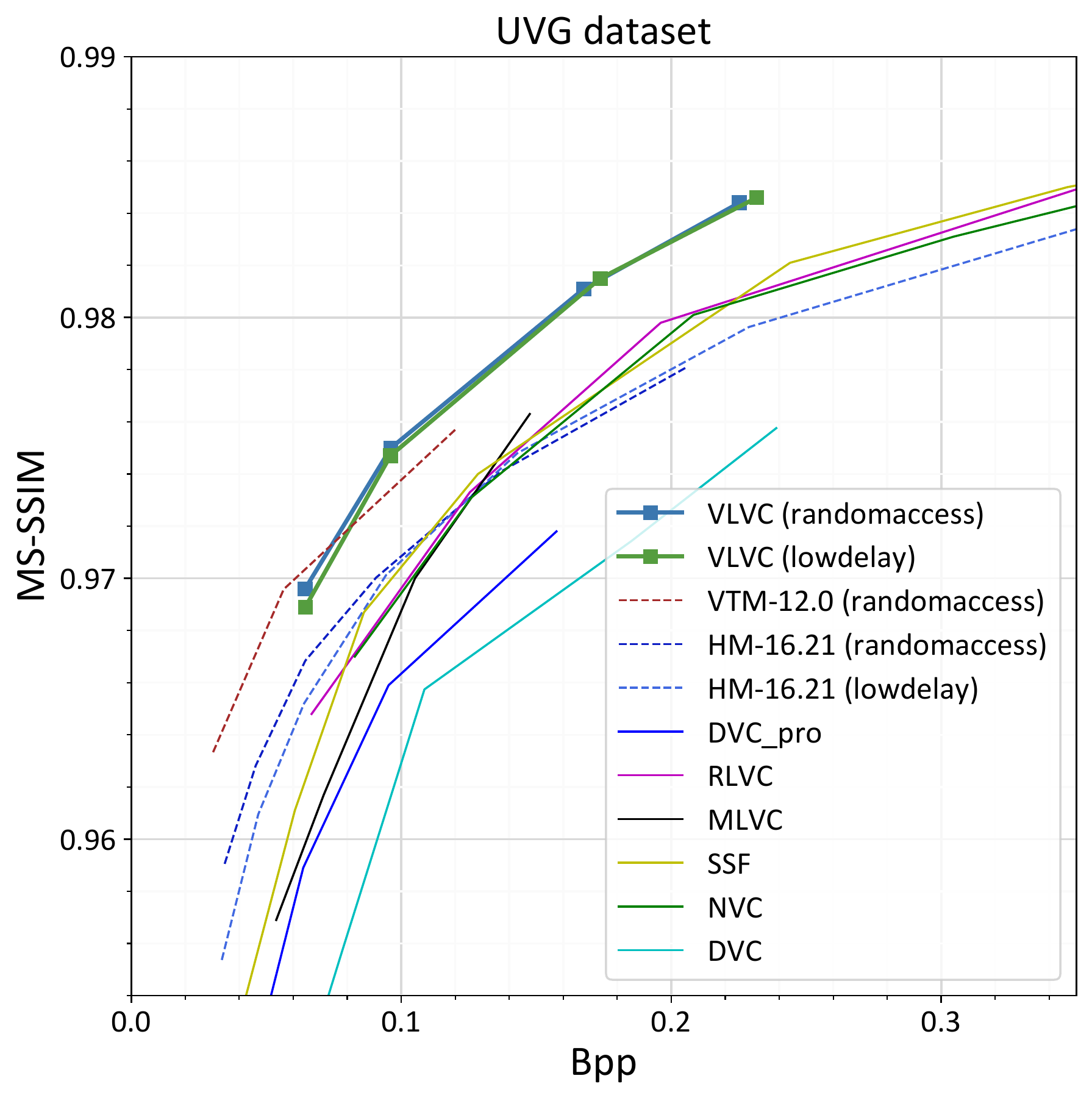}
}\hfill
\subfloat{
    \includegraphics[width=0.32\linewidth]{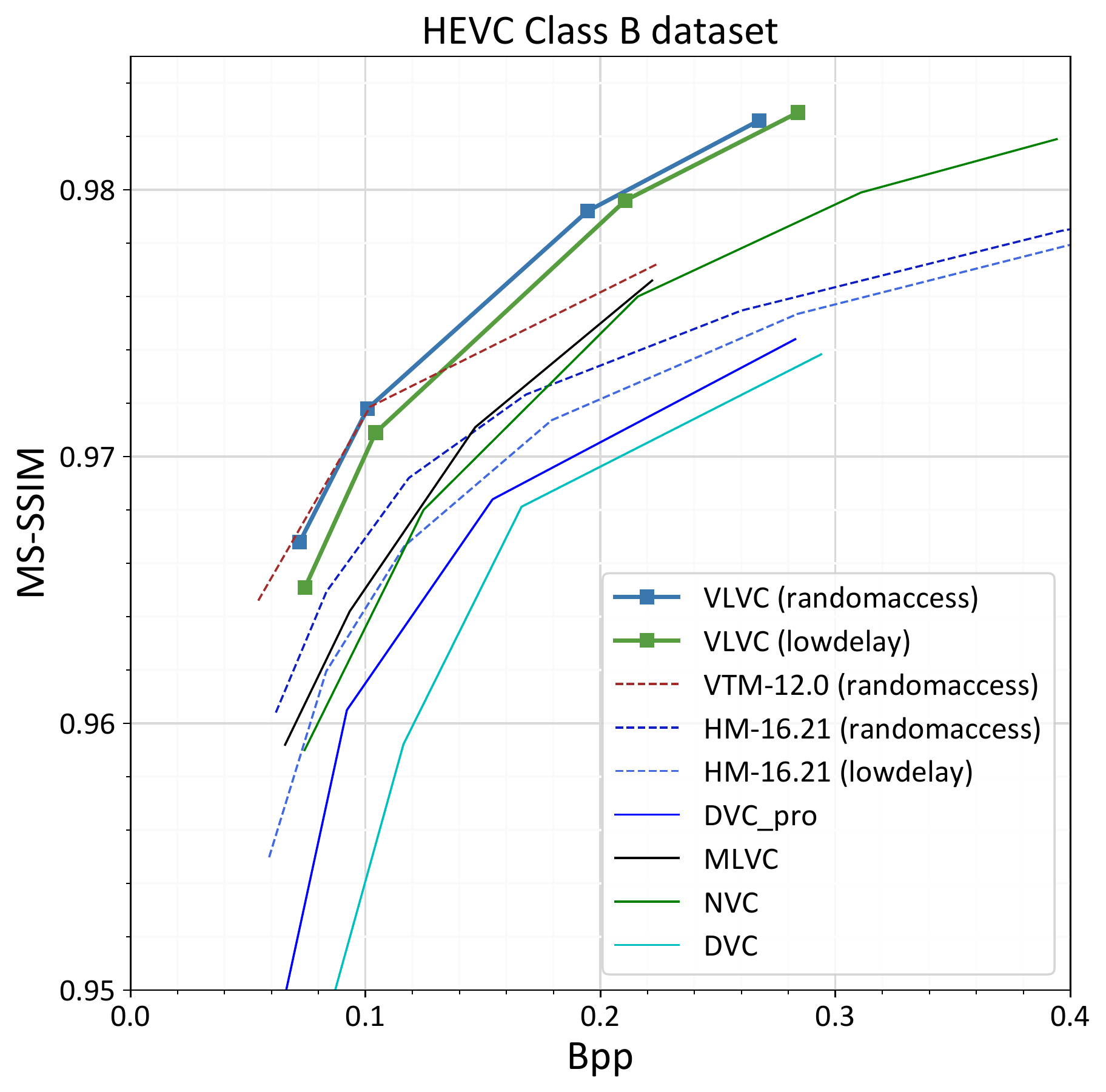}
}\hfill
\subfloat{
    \includegraphics[width=0.323\linewidth]{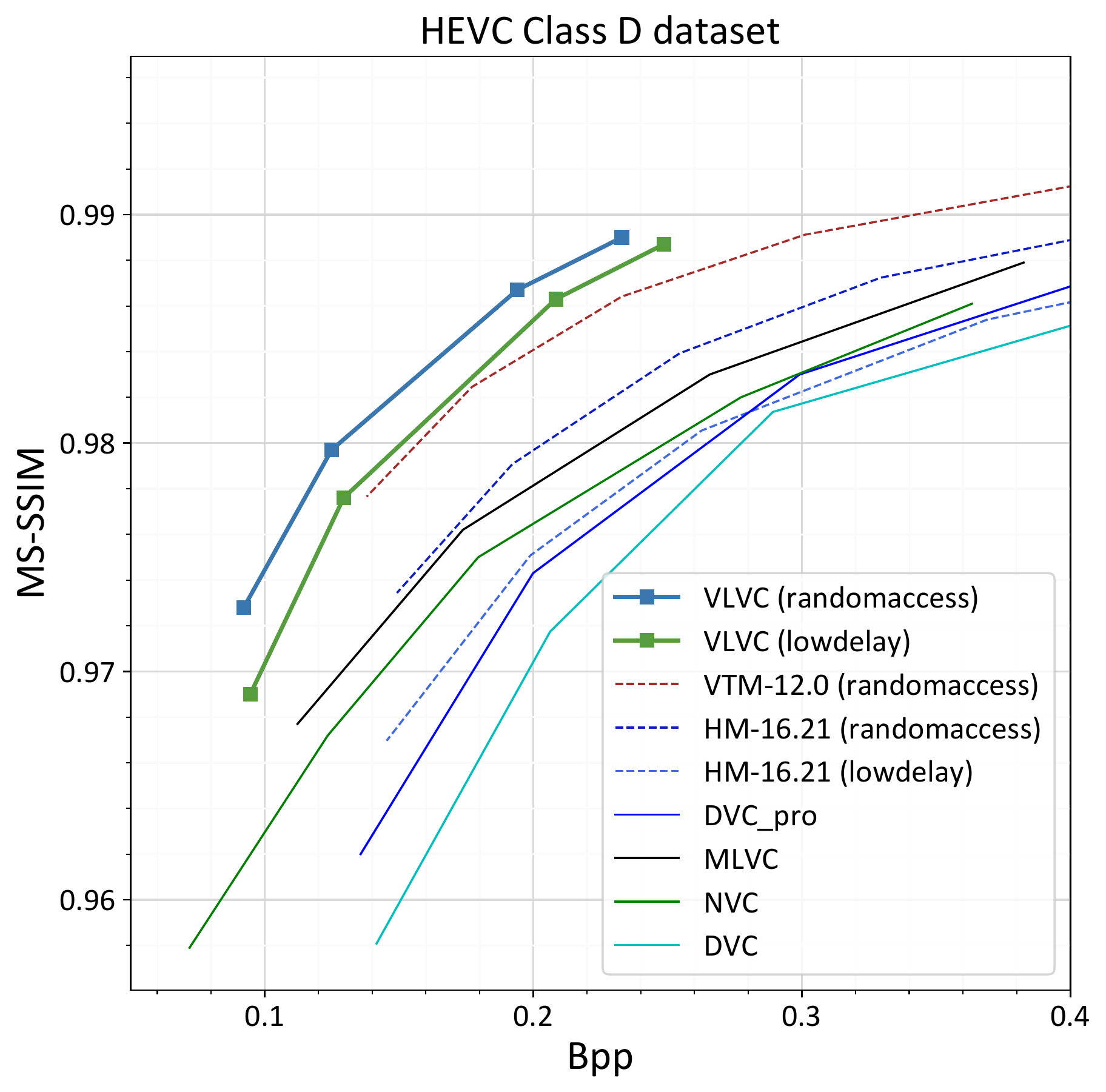}
}
\caption{Rate-distortion curves on UVG, HEVC ClassB and HEVC ClassD datasets. Top row: PSNR performance. Bottom row: MS-SSIM performance.}
\label{fig:performance}
\end{figure*}

\begin{figure*}[htb]
\centering
\subfloat[\label{arch-1a}]{
    \includegraphics[angle=270, width=\linewidth]{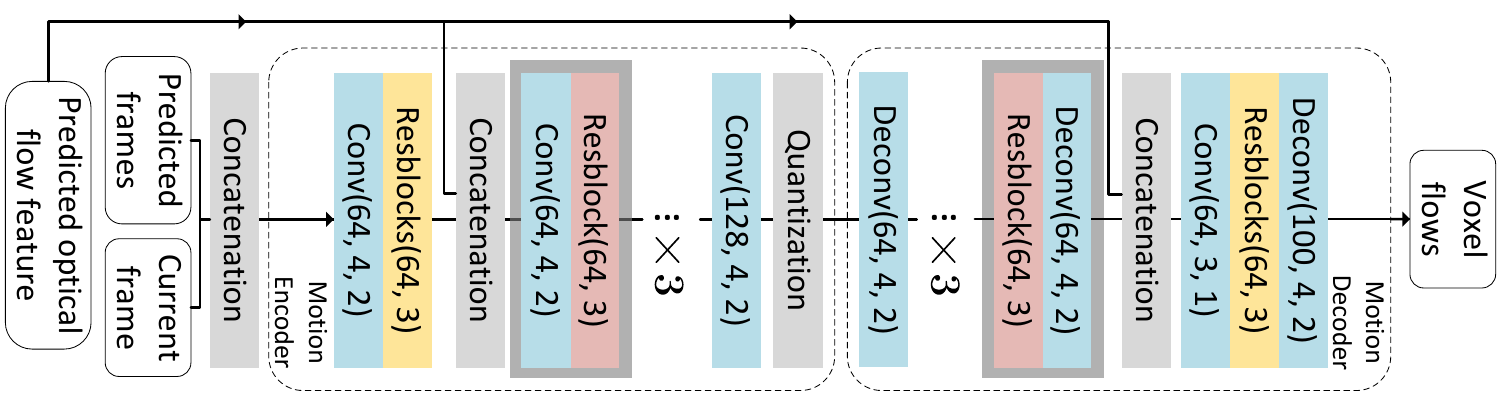}
}\\
\subfloat[\label{arch-1b}]{
    \includegraphics[angle=270, width=\linewidth]{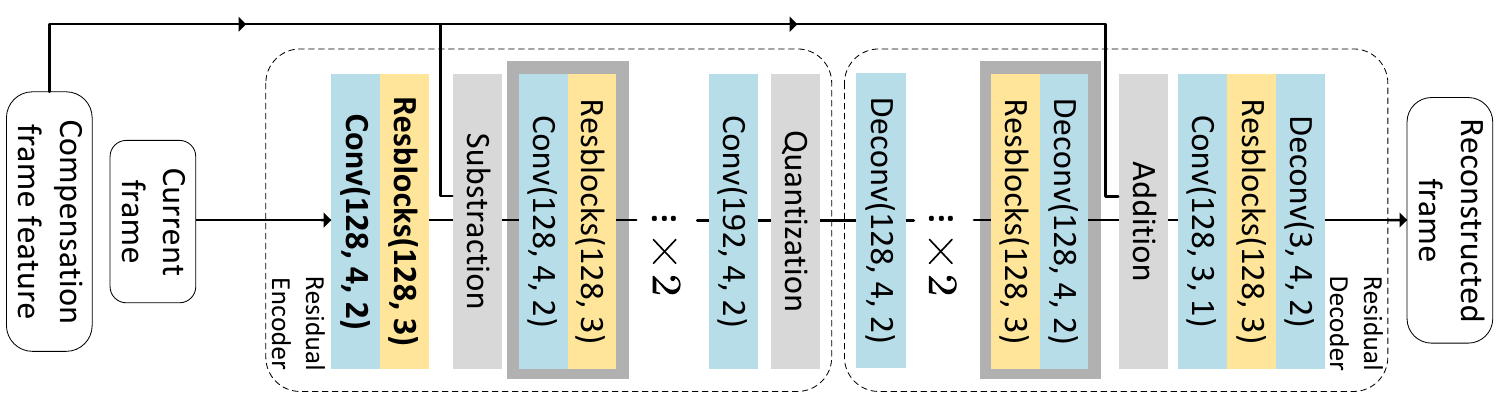}
}
\caption{Detailed structure of (a) the Motion Encoder/Decoder and (b) the Residual Encoder/Decoder. \textit{Conv(C, K, S)} and \textit{Deconv(C, K, S)} represent the convolution and deconvolution layers with C output channels and a kernel size of K and a stride of S. The predicted optical flow feature in Fig.~\ref{arch-1a} is the output of \textit{Optical Flow Encoder}, and the compensation frame feature in Fig.~\ref{arch-1b} is generated by transforming the compensation frame with the first two layers (
bold font layers) of \textit{Residual Encoder} .
The details of \textit{Optical Flow Encoder}, \textit{Resblock} and \textit{Resblocks} are shown in Fig.~\ref{fig:arch-2}.}
\label{fig:arch-1}
\end{figure*}

\paragraph{Flow reversal via softmax splatting}
While the forward flow filed $\boldsymbol{f}_{t_j \rightarrow t}$ can be calculated by the per-pixel polynomial functions, it cannot be directly used for backward warping. Although some previous compression works such as \cite{yang2020learning} directly inverse the forward flow as the backward flow, it is inaccurate and may encounter some issues such as inconsistent occlusion \cite{niklaus2020softmax,xu2019quadratic}. Therefore, we adopt a flow reversal layer to generate the backward flow $\boldsymbol{f}_{t \rightarrow t_j}$ by softmax splatting \cite{niklaus2020softmax}:

\begin{equation}\label{eq:3}
 \centering
\begin{aligned}
\boldsymbol{f}_{t \rightarrow t_j} = \frac{\sum\limits^{\rightarrow}{( \exp(\boldsymbol{Z})\cdot(-\boldsymbol{f}_{t_j \rightarrow t}),\boldsymbol{f}_{t_j \rightarrow t})}}{\sum\limits^{\rightarrow}{( \exp(\boldsymbol{Z}), \boldsymbol{f}_{t_j \rightarrow t})}},
\end{aligned}
\end{equation}
where $\sum\limits^{\rightarrow}$ is the summation splatting defined in \cite{niklaus2020softmax}, and $\boldsymbol{Z}$ is an importance mask generated as:

\begin{equation}\label{eq:4}
\begin{aligned}
\boldsymbol{Z} = q(\boldsymbol{\hat{x}}_{t_j}, -\frac{1}{k}\sum_{i=1}^{k}{\Vert \boldsymbol{\hat{x}}_{t_j} -  \mathop{w}\limits^{\leftarrow}(\boldsymbol{\hat{x}}_{t_i}, \boldsymbol{f}_{t_j \rightarrow t_i})\Vert_1}).
\end{aligned}
\end{equation}
Here $q$ is a small network and $\mathop{w}\limits ^{\leftarrow}$ is the bilinear backward warping operator. With this softmax splatting process, we finally obtain the predicted backward optical flow, which is then used to generate the predicted frames as well as the auxiliary information for motion encoding and decoding.

\textbf{Discussion.} Our proposed polynomial function is a mathematical formulation for different motion modes conforming to physics, where the first-order and second-order coefficients can be interpreted as the speed and the acceleration of motion, respectively. This physical interpretation comes from recent work~\cite{xu2019quadratic} for video interpolation, and the effectiveness of quadratic/cubic motion modeling is experimentally verified in previous works~\cite{xu2019quadratic, chi2020all}.
In this paper, it is original to propose such a unified form for arbitrary reference structures.
The polynomial function cannot be replaced with a fixed linear/quadratic function since it is bound with the coding flexibility of our method. For example, if choosing linear function as an alternative, it only describes the motion according to two reference frames and cannot handle more reference frames.

\subsection{Loss function}
In previous works, the reference frames are determined according to pre-defined prediction modes. For example, the work of \cite{lin2020m} applies four unidirectional reference frames, where the reference set is $\{\boldsymbol{\hat{x}}_{t-4}, \boldsymbol{\hat{x}}_{t-3}, \boldsymbol{\hat{x}}_{t-2}, \boldsymbol{\hat{x}}_{t-1}\}$. The work of \cite{djelouah2019neural} applies $\{(\boldsymbol{\hat{x}}_{t-1}, \boldsymbol{\hat{x}}_{t+1}), (\boldsymbol{\hat{x}}_{t-2}, \boldsymbol{\hat{x}}_{t+2}), (\boldsymbol{\hat{x}}_{t-3}, \boldsymbol{\hat{x}}_{t+3})\}$ as the reference set for bilinear prediction. In this paper, to optimize a versatile compression model, the model will have access to various reference structures during training to adapt to different prediction modes. Therefore, we apply the loss function to cover the frames in the entire GOP as:

\begin{equation}\label{eq:GOP-loss-ours}
\centering
\begin{aligned}
\mathcal{L}
= 
\frac{1}{T} \sum_{t=2}^T &[  R(\boldsymbol{\hat{m}}_t, \boldsymbol{\hat{r}}_t|\boldsymbol{\hat{x}}_{t_i}, ...,  \boldsymbol{\hat{x}}_{t_j}  ) \\
&+\lambda \cdot  d(\boldsymbol{x}_{t}, \boldsymbol{\hat{x}}_{t} |\boldsymbol{\hat{x}}_{t_i}, ...,  \boldsymbol{\hat{x}}_{t_j} )  ] .
\end{aligned}
\end{equation}
Here, $T$ is the clip length, which is set to 7 during training. $\{\boldsymbol{\hat{x}}_{t_i}, ...,  \boldsymbol{\hat{x}}_{t_j} \}$ represents different reference sets that may vary with different $t$ and mini-batches. $R(\boldsymbol{\hat{m}}_t, \boldsymbol{\hat{r}}_t)$ is the rate of motion and residual. $d(*,*)$ is the distortion metric. For simplicity, we omit to write down the rate-distortion terms of intra frame ($R_1 + \lambda \cdot d(\boldsymbol{x}_{1}, \boldsymbol{\hat{x}}_{1})$) in this loss function.

\begin{figure*}[htb]
\centering
\subfloat[\label{fig:arch-2a}]{
    \includegraphics[width=0.27\linewidth]{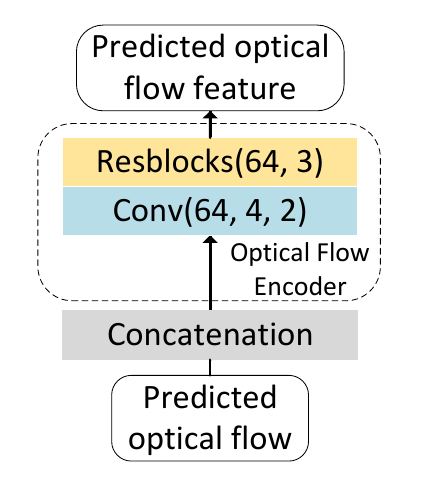}
}\hfill
\subfloat[\label{fig:arch-2b}]{
    \includegraphics[width=0.27\linewidth]{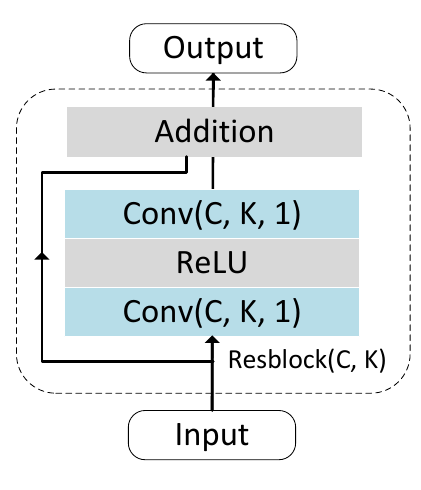}
}\hfill
\subfloat[\label{fig:arch-2c}]{
    \includegraphics[width=0.27\linewidth]{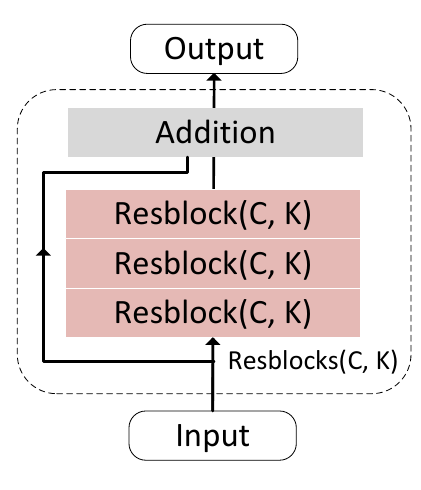}
}
\caption{Detailed structure of \textit{Optical Flow Encoder}, \textit{Resblock} and \textit{Resblocks}. \textit{Conv(C, K, S)} represents the convolution layer with C output channels and a kernel size of K and a stride of S.}
\label{fig:arch-2}
\end{figure*}

\section{Experiments}\label{sec:experiments}
\subsection{Experimental setup}
\paragraph{Model details}
The model structure details are provided in Fig.~\ref{fig:arch-1} and Fig.~\ref{fig:arch-2}. The motion/residual compression modules are two autoencoder-based networks, where the bit-rate is estimated by the factorized and hyperprior model~\cite{balle2018variational, minnen2018joint}, respectively. 
We employ the off-the-shelf PWC-Net \cite{sun2018pwc} as the optical flow estimation network only in our generalized flow prediction module. 
We employ feature residual coding~\cite{feng2020learned} instead of pixel residual coding for better performance. Our intra codec is also an autoencoder-based network.

\paragraph{Training and testing sets}
The models were trained on the Vimeo-90k septuplets dataset~\cite{xue2019video} which consists of 89800 video clips with diverse content. The video clips are randomly cropped to 128 $\times$ 128 or 256 $\times$ 256 pixel for training.
The HEVC common test sequences~\cite{HM}, UVG dataset~\cite{mercat2020uvg} and MCL-JCV dataset~\cite{wang2016mcl}are used for evaluation. The HEVC Classes B,C,D and E contain 16 videos with different resolution and content. The UVG dataset contains seven 1080p HD video sequences with 3900 frames in total.

\paragraph{Implementation details}
 We optimize five models for MSE and four models for MS-SSIM~\cite{wang2003multiscale}. 
 We use the Adam optimizer~\cite{kingma2014adam} with a batch size of 8 and an initial learning rate of $5 \times 10^{-5}$. It is difficult to stably train the whole model from scratch. We first separately pre-train the intra-frame coding models and inter-frame coding models for MSE, with $128 \times 128$ video crops and 1,200,000 training steps. Then we jointly optimize both the models with the loss Eq.~(\ref{eq:GOP-loss-ours}) for 100,000 steps using different metrics and $\lambda$ values. Finally, we fine-tuning all the models for $20,000$ steps with a crop size of $256 \times 256$ and a reduced learning rate of $1 \times 10^{-5}$.

\paragraph{Evaluation Setting} 
We measure the quality of reconstructed frames using  PSNR and MS-SSIM~\cite{wang2003multiscale} in the RGB colorspace. The bit per pixel (bpp) is used to measure the average number of bits. We compare our method with the traditional video coding standards H.265/HEVC and H.266/VVC, as well as the state-of-art learning-based methods including SSF \cite{Agustsson_2020_CVPR}, DVC\_pro \cite{lu2020end}, MLVC \cite{lin2020m}, NVC \cite{liu2020neural} and RLVC \cite{yang2020learning}.
Recent works for learned video compression usually evaluate H.265 by using FFmpeg, which performance is much lower than official implementation. In this paper, we evaluate H.265 and H.266 by using the implementation of the standard reference software HM 16.21~\cite{HM} and VTM 12.0~\cite{VTM}, respectively. We use the default low delay and random access configurations, and modify the GOP structures for a fair comparison. Detailed configurations can be found in the Appendix.

\begin{table*}[htb]
\centering
\resizebox{\linewidth}{!}{
\tiny
\begin{tabular}{c l r r r r r r}
\hline
Metric & Codec & UVG & MCL-JCV & ClassB & ClassC & ClassD & ClassE \\ \hline
\multirow{6}*{MS-SSIM} & VVC \cite{vvc} & -0.97\%  & - & -4.71\% & -7.37\% & -18.25\% & -6.31\% \\ 
& SSF \cite{Agustsson_2020_CVPR} & -28.94\% & -23.74\% & - & - & - & - \\ 
& MLVC \cite{lin2020m} & -33.02\% & - & -35.11\% & -45.56\% & -41.48\% & -46.31\% \\ 
& DVC\_pro \cite{lu2020end} & -51.03\% & - & -47.58\% & -45.16\% & -50.25\% & -31.99\% \\ 
& NVC \cite{liu2020neural} & -31.34\% & - & -36.59\% & -42.86\% & -45.86\% & -24.66\% \\ 
& RLVC \cite{yang2020learning} & -29.12\% & -32.35\% & - & - & - & - \\ \hline

\multirow{6}*{PSNR} & Djelouah~et~al. \cite{djelouah2019neural} & -8.52\% & -33.41\% & - & - & - & - \\ 
& SSF \cite{Agustsson_2020_CVPR} & -31.27\% & -20.29\% & - & - & - & - \\ 
& MLVC \cite{lin2020m} & -29.42\% & -40.25\% & -19.63\% & -26.70\% & -17.77\% & 6.42\% \\ 
& DVC\_pro \cite{lu2020end} & -24.16\% & - & -34.49\% & -14.30\% & -21.64\% & -3.94\% \\ 
& NVC \cite{liu2020neural} & -31.93\% & - & -34.97\% & -23.84\% & -32.15\% & -3.59\% \\ 
& RLVC \cite{yang2020learning} & -23.89\% & -24.11\% & - & - & - & - \\ \hline
\end{tabular}
}
\vspace{4pt}
\caption{The BD-rate savings of VLVC relative to other video codecs in terms of PSNR and MS-SSIM on different datasets.}
\label{table:BDrate}
\end{table*}

\subsection{Performance}
Fig.~\ref{fig:performance} shows the rate-distortion curves. It can be observed that our proposed method outperforms existing learned video compression methods in both PSNR and MS-SSIM.
Most importantly, our model is the first end-to-end learned video compression method that outperforms the latest VVC/H.266 standard reference software in terms of MS-SSIM. Note that the``VLVC (randomaccess)'' and ``VLVC (lowdelay)'' are two different coding configurations from the same model. 
 
In Table~\ref{table:BDrate}, we provide the BD-rate~\cite{Bjntegaard2001CalculationOA} savings of VLVC (randomaccess) relative to other video codecs in terms of both PSNR and MS-SSIM, where negative values indicate BD-rate savings. Our proposed VLVC saves more bit-rate on various benchmark datasets.

\subsection{Ablation Study and Analysis}
\begin{figure*}[htb]\label{fig:ablation}
\centering
\subfloat[\label{fig:ablation-a}]{
    \includegraphics[width=0.32\linewidth]{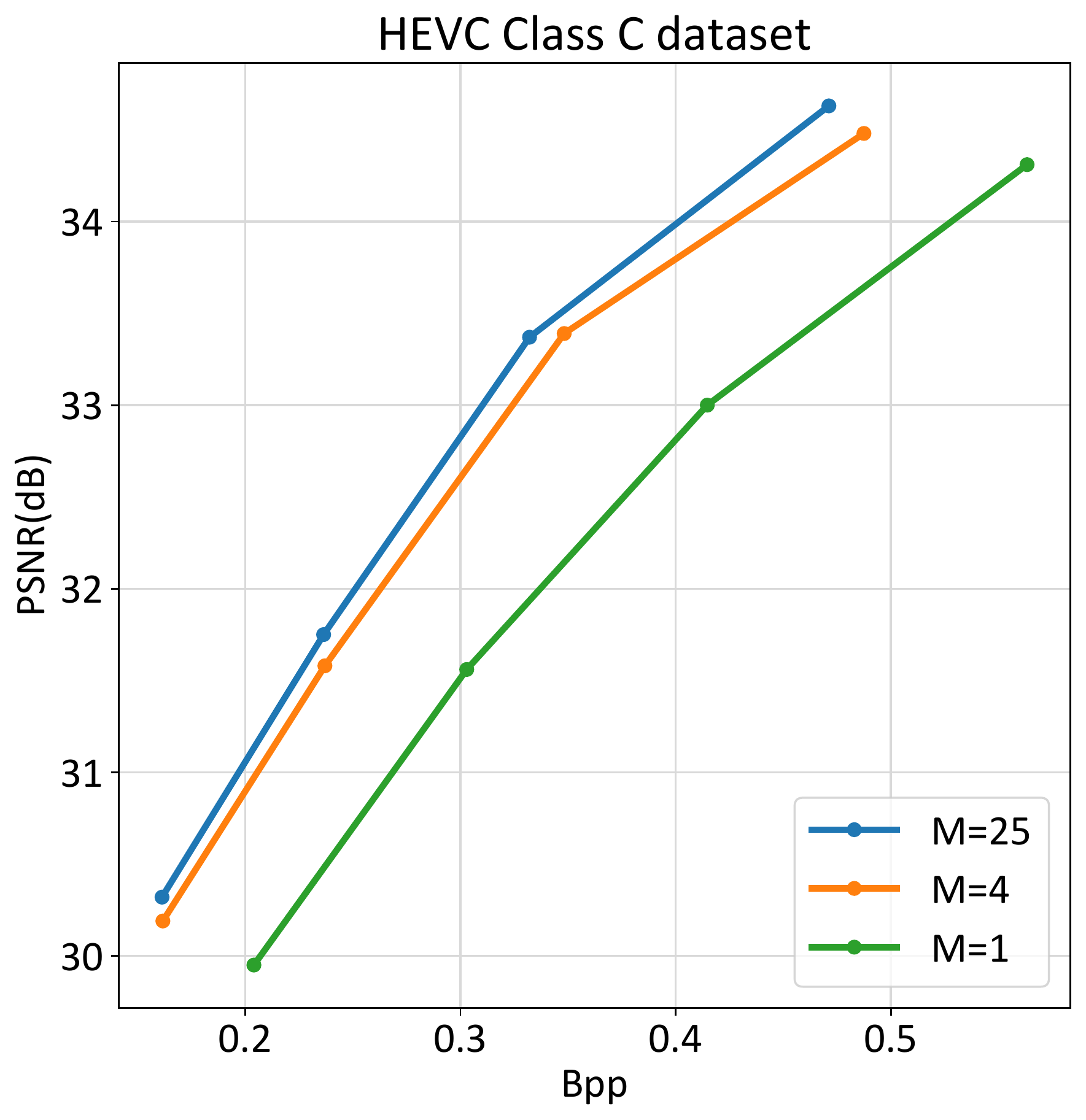}
}\hfill
\subfloat[\label{fig:ablation-b}]{
    \includegraphics[width=0.315\linewidth]{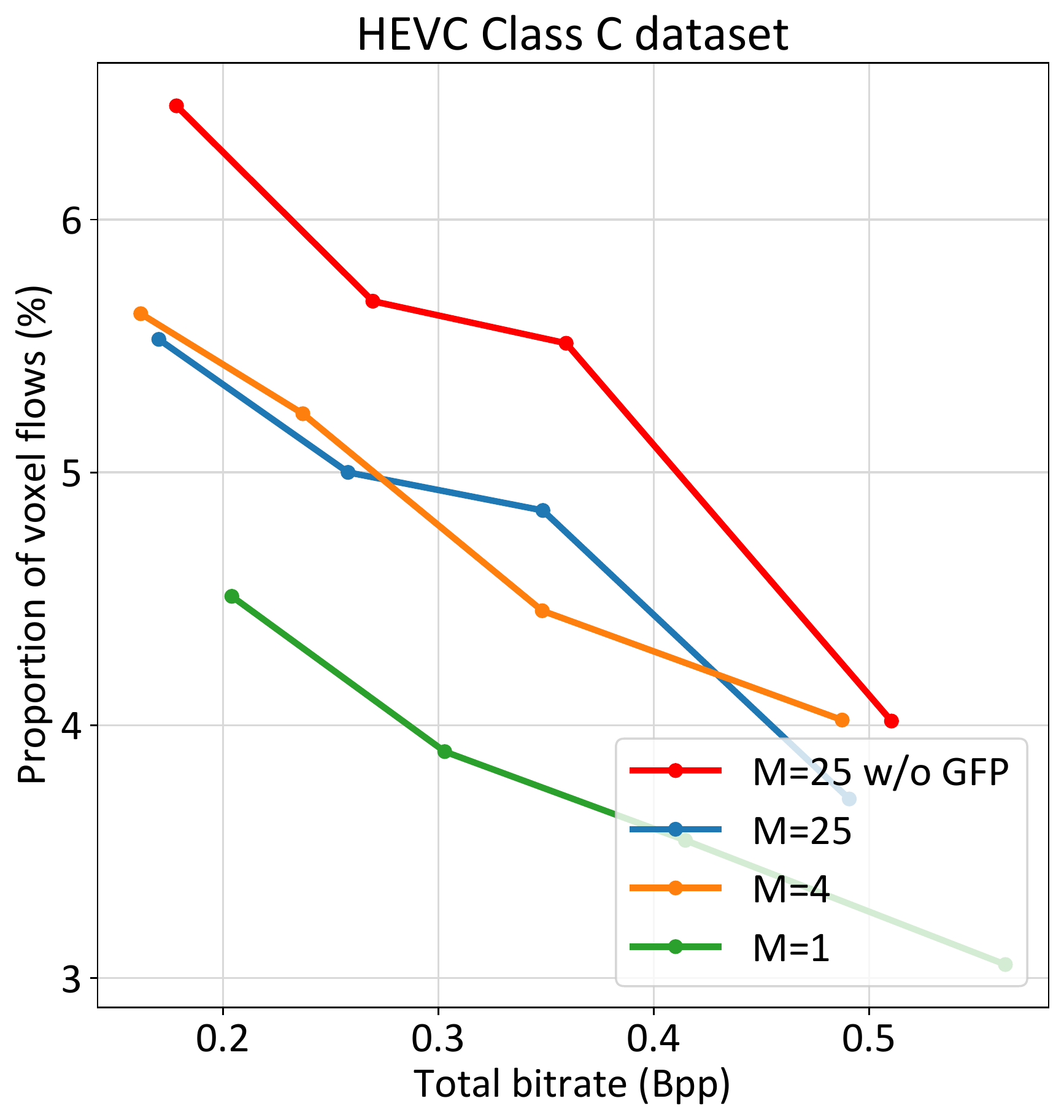}
}\hfill
\subfloat[\label{fig:ablation-c}]{
    \includegraphics[width=0.32\linewidth]{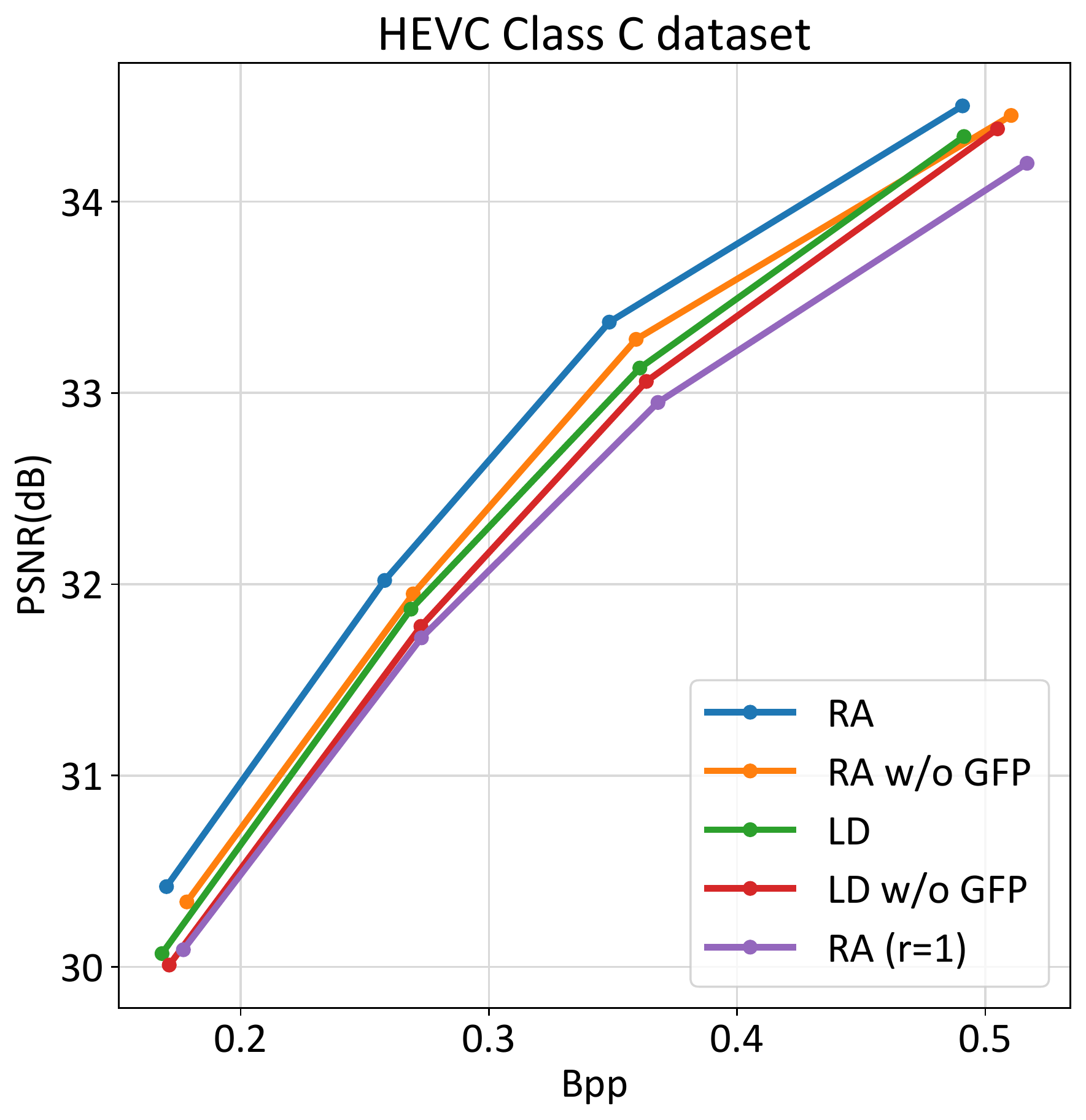}
}
\caption{Ablation studies on: (a) the number of voxel flows, (b) the proportion of voxel flows in total bitrate and (c) the different coding configurations.}
\end{figure*}

\paragraph{The effect of the voxel flow number}
As shown in Fig.~\ref{fig:ablation-a}, the number $M$ of voxel flows significantly influences the overall rate-distortion performance. More voxel flows enable our codec to better model motion uncertainty. Our proposed weighted warping with multiple voxel flows achieves about 1dB gain compared with the conventional trilinear warping with a single voxel flow. 
Note that the performance gain is nearly saturated for $M=25$, which is used as the default value in our models. 

We also investigate the additional bitrate cost of multiple voxel flows. As shown in Fig.~\ref{fig:ablation-b}, the proportion of multiple voxel flows in the total bitrate of video coding increases about $\frac{1}{3}$ at the same bitrate. In other words, our model can learn to improve the overall compression performance by transmitting a proper amount of voxel flows.

\paragraph{Versatile coding configurations}
The proposed methods can deal with various prediction modes. To evaluate the effectiveness of coding flexibility as well as the effectiveness of the proposed generalized flow prediction module, we simply change the input coding configurations of the same trained models. Random access (bidirectional reference) and low delay (unidirectional reference) coding modes are denoted as ``RA'' and ``LD'', respectively. As shown in Fig.~\ref{fig:ablation-c}, the ``RA'' mode achieves a compression gain of about 0.4dB, compared with the ``LD'' mode. Furthermore, the performance dropped about 0.1dB \textasciitilde 0.3dB when we turn off the GFP block for different coding settings, noted as ``w/o GFP''. We also illustrate the bitrate reduction of the voxel flows shown in Fig.~\ref{fig:ablation-b}, where the model of ``M=25'' reduces the bitrate of voxel flows by about $\frac{1}{6}$ compared to the model of ``M=25 w/o GFP''. Finally, we change the number of the reference frames for weighted  warping, which is set to 2 as default. We reduce the number to 1 in ``RA'' mode, noted as ``RA (r=1)'', which performance is even worse than ``LD'' mode.

\begin{figure*}[htb]
\centering
\subfloat[\label{fig:visflow-1}]{
    \includegraphics[width=0.97\linewidth]{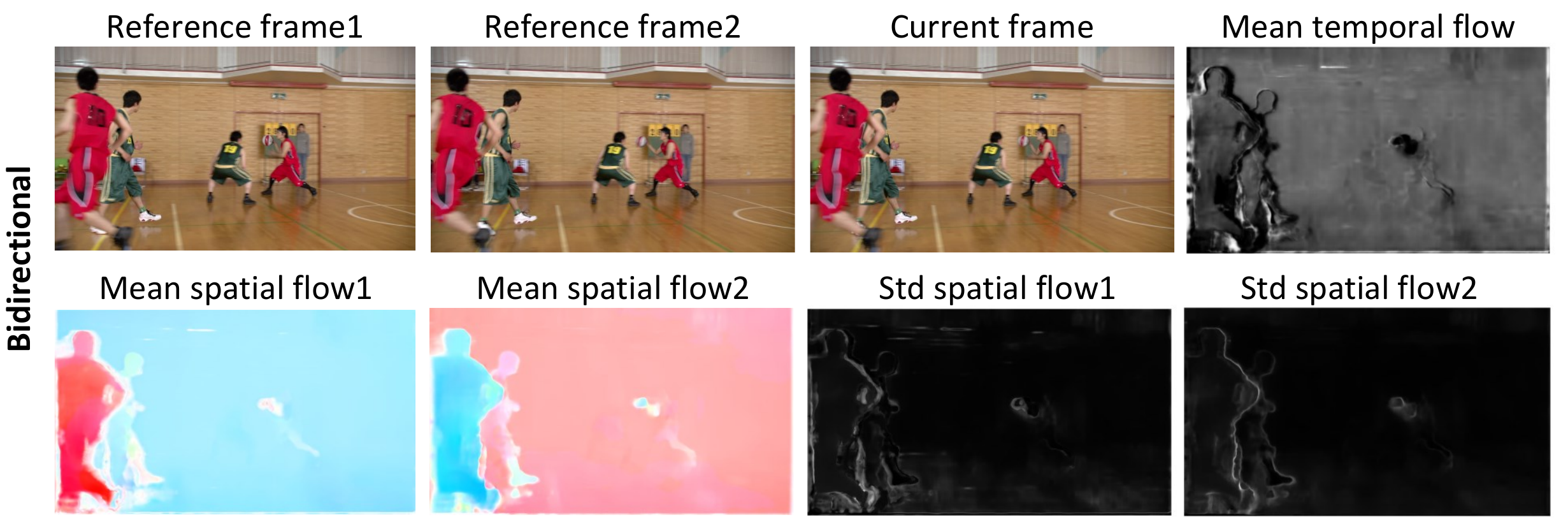}
}\\
\subfloat[\label{fig:visflow-2}]{
    \includegraphics[width=0.97\linewidth]{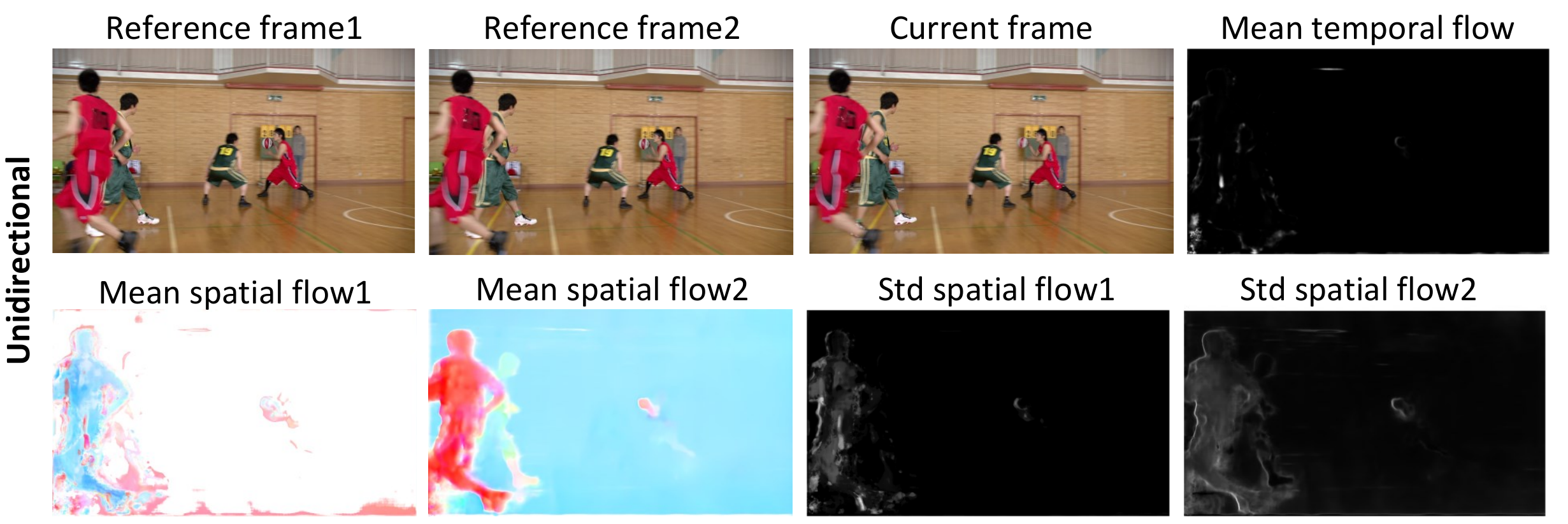}
}
\caption{Visualization of the voxel flows for the same target frame with different reference frames, generated by the same model. (a) Bidirectional reference. (b) Unidirectional reference. For the ``Mean spatial flow'', like optical flow, different color represents different motion direction. For ``Std spatial flow'', different colors represent the difference among multiple flows, whereas the white color represents a high std value. For ``Mean temporal flow'', different color means different temporal reference position, where the white color means  $\boldsymbol{g}_z$ is close to reference 1 and the black color means $\boldsymbol{g}_z$ is close to reference 2.}
\label{fig_ap:visflow}
\end{figure*}

\paragraph{Visualization of voxel flows}
The proposed voxel flows contain multiple 3-channel voxel flows $\{(\boldsymbol{g}^i_x, \boldsymbol{g}^i_y, \boldsymbol{g}^i_z)\}_{i=1}^M$ and their weights $\{\boldsymbol{g}^i_w\}_{i=1}^M$. 
We separately visualize the weighted temporal and spatial flow maps. The mean temporal flow map $\boldsymbol{\bar{g}}_z = \sum_i{\boldsymbol{g}^i_w \cdot \boldsymbol{g}^i_z}$ describes the weighted centroid of voxel flows along the time axis. As shown in the fourth column of Fig.~\ref{fig:visflow-1}, the $\boldsymbol{\bar{g}}_z$ performs like an occlusion map for bidirectional frame prediction. The pixels in the black area (e.g. background around the basketball players) are covered in the first reference frame. 
Therefore, the voxel flows of the black area pay more attention to the second reference. 
For the unidirectional frame prediction shown in Fig.~\ref{fig:visflow-2}, however, the $\boldsymbol{\bar{g}}_z$ generated by the same model is almost black everywhere, demonstrating the flexibility of voxel flows for different reference structures. 

We also visualize the weighted mean and weighted standard deviation of spatial flow maps (noted as ``Mean spatial flow'' and ``Std spatial flow'') to investigate the spatial distribution of voxel flows. We group the voxel flows to the nearest reference according to $\boldsymbol{\bar{g}}_z$. As shown in Fig.~\ref{fig_ap:visflow}, the spatial mean of grouped voxel flows has similar distribution with optical flow. The voxel flows have large variance in the area of motion, occlusion and blur, shown in the ``Std spatial flow''. Single optical flow sometimes cannot find an accurate reference pixel (e.g. occlusion area). Multiple flow weighted warping can model the uncertainty of flow using multiple reference pixels, yielding better performance. 

The scale-space warping \cite{Agustsson_2020_CVPR} also models the uncertainty of flow using different Gaussian smoothed reference values, where the kernel weights and shape are fixed. In this paper, weighted warping allows the model to freely learn the shape and weights of the 3D ``smoothing kernel'', which is more flexible and generalized.

\begin{table}[htb]
\centering
\resizebox{\linewidth}{!}{
\begin{tabular}{p{42pt}p{20pt}p{20pt}p{20pt}p{20pt}p{20pt}}
\hline
             & DVC & VLVC (LDP) & MLVC & VLVC (LDB) & VLVC (RA) \\ \hline
encoding (s) & 0.59	& 0.46 & 1.23 & 0.86 & 0.86 \\ 
decoding (s) & 0.32 & 0.33 & 0.99 & 0.74 & 0.74 \\ \hline
\end{tabular}
}
\vspace{4pt}
\caption{Encoding and decoding time of different codecs.}
\label{table:runtime}
\end{table}

\subsection{Model Complexity.}
We evaluate the encoding/decoding time with one 2080TI GPU (11GB memory) and one Intel(R) Xeon(R) Gold 5118 CPU @ 2.30GHz. The runtime of VLVC is comparable with recent learning-based codecs, such as DVC~\cite{lu2019dvc} (single reference frame) and MLVC~\cite{lin2020m} (multiple reference frames). For a fair comparison, we reimplement the works of DVC and MLVC using PyTorch and compare the network inference time on 1080p videos, except for the time of arithmetic coding (on CPU).

As shown in Table~\ref{table:runtime}, the VLVC (LDP), VLVC (LDB) and VLVC (RA) are low delay P (unidirectional, single reference frame), low delay B (unidirectional, multiple reference frames) and random access (bidirectional, multiple reference frames) modes of VLVC, respectively. The runtime of our method is slightly less than DVC and MLVC under similar coding configurations. The time of arithmetic coding is not included for comparison because it is a common part of any codec and is sensitive to implementation. During the test, the implementation of this part is commonly off-the-shelf where different compression models can use the same one. For VLVC with arithmetic coding, the overall coding speed of RA mode is about 0.7 fps. Besides, our proposed weighted voxel flow based warping takes about 0.033s per frame for HD 1080.

The total size of our inter-frame compression model is about 70MB (except for the off-the-shelf optical flow estimation network PWC-Net\cite{sun2018pwc}). Our model size is numerically large since we apply three residual blocks after each downsampling/upsampling layer to enhance the motion/residual compression network, compared with DVC (about 11MB). This is a trivial enhancement that leaves a large room for model slimming.

The training of VLVC consists of three parts: I-frame codec pretraining (1 day on a 2080Ti GPU), inter-frame codec pretraining (2 days on a 2080Ti GPU) and joint training (2 days on four 2080Ti GPUs). The training time is comparable with recent works like SSF \cite{Agustsson_2020_CVPR} (4 days on a NVidia V100 GPU) and DVC\_pro \cite{lu2020end} (4 days on two GTX 1080Ti GPUs).

\textbf{Discussions.} In case of single-reference prediction, our model is faster than DVC, since the flow prediction module is turned off in this case. And our model does not require an optical flow extractor in the motion encoder. In case of multiple-reference prediction, the flow prediction in our model is accomplished by solving a multi-variant equation, different from MLVC which applies a complex flow fusion module. Therefore, our model is also faster than MLVC in the scenario of prediction with multiple reference frames. 

\subsection{Subjective Comparison}
\begin{figure*}[htb]
\centering
\subfloat[\label{fig:vis1}]{
    \includegraphics[width=0.49\linewidth]{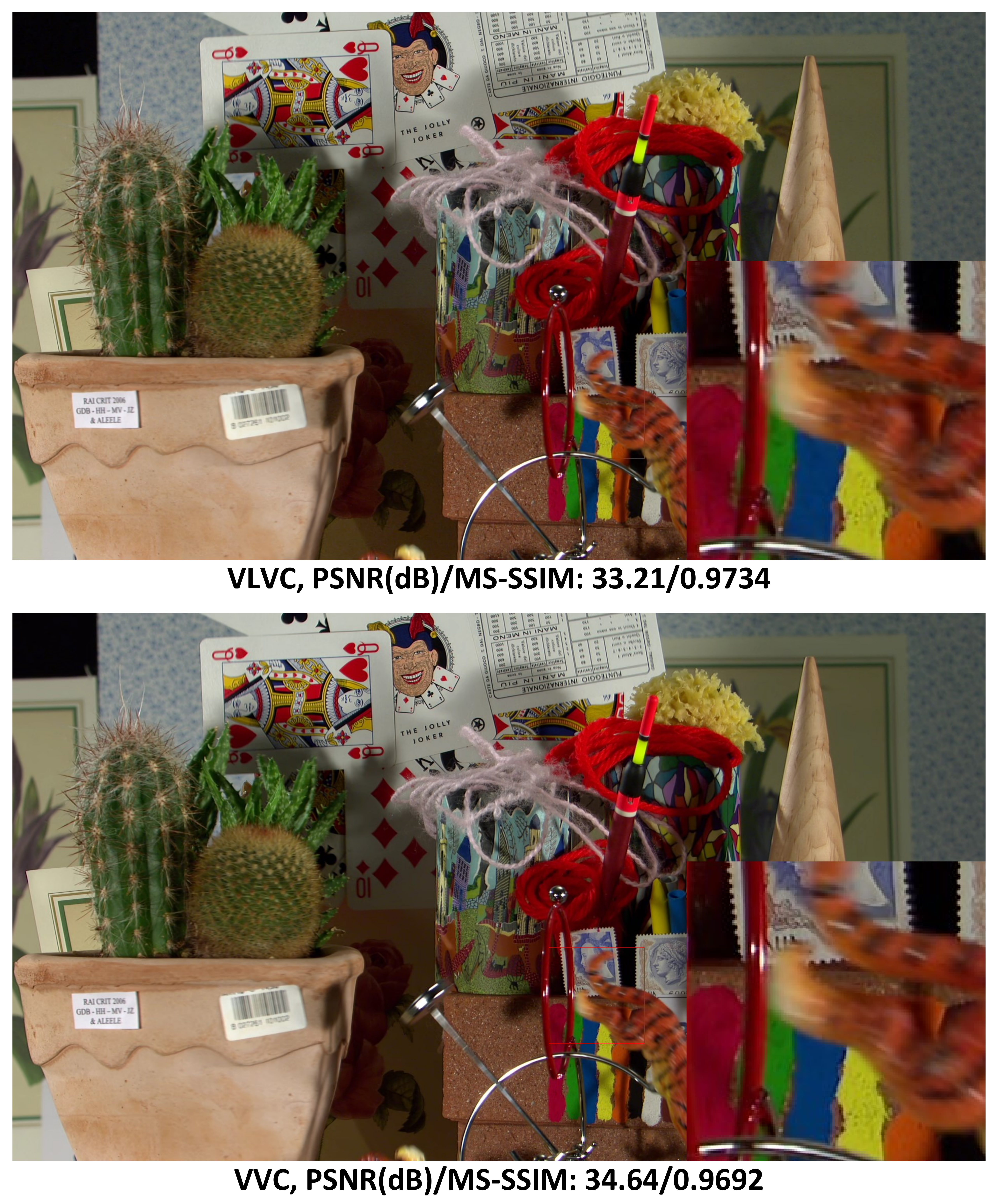}
}
\subfloat[\label{fig:vis2}]{
    \includegraphics[width=0.49\linewidth]{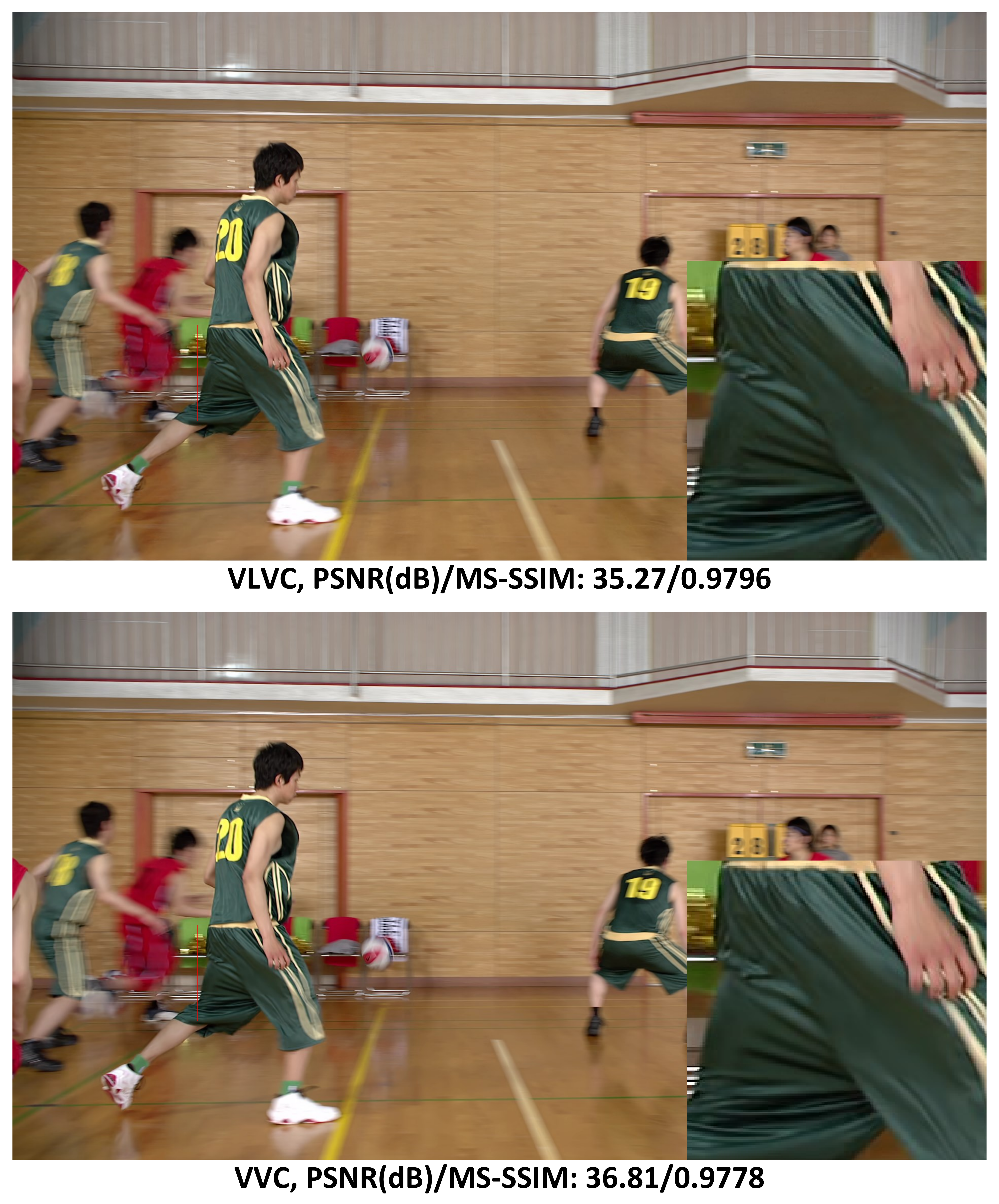}
}
\caption{Subjective comparison between our proposed VLVC and VVC on reconstructed frames of the videos ‘Cactus‘ (left column) and ‘BasketballDrive‘ (right column) in HEVC ClassB. The reconstructed frames of VLVC are sharper and richer in texture while the average bpp is smaller.}
\label{fig:vis}
\end{figure*}

To verify if high MS-SSIM scores lead to high subjective quality in our models, we visualize the reconstruction of VLVC and VVC with similar average bitrate on the HEVC ClassB dataset (0.1945 bpp and 0.2238 bpp, respectively). As shown in Fig.~\ref{fig:vis1} and Fig.~\ref{fig:vis2}, compared with VVC, the VLVC's reconstructed frames with higher MS-SSIM scores are sharper and richer in texture and has better subjective quality, while the corresponding PSNR values are lower.

\section{Conclusion}\label{sec:conclusion}
In this paper, we propose a versatile learned video coding (VLVC) framework that allows us to train one model to support various inter prediction modes. To this end, we apply voxel flows as a motion information descriptor along both spatial and temporal dimensions. The target frame is then predicted with the proposed weighted trilinear warping using multiple voxel flows for more effective motion compensation. Through formulating various inter prediction modes by a unified polynomial function, we design a novel flow prediction module to predict accurate motion trajectories. In this way, we significantly reduce the bit cost of encoding motion information. Thanks to above novel motion compensation and flow prediction, VLVC not only achieves the support of different inter prediction modes but also yields competitive R-D performance compared to conventional VVC standard, which fosters practical applications of learned video compression technologies.    

\section*{Acknowledgements}
This work was supported in part by NSFC under Grant U1908209, 61632001, 62021001 and the National Key Research and Development Program of China 2018AAA0101400.

\appendix
\section*{Appendix: Configurations of the HEVC/VVC reference software}
Most of recent learning-based video codecs are evaluated in sRGB color space. To make a fair comparision, we first convert the source video frames from YUV420 to RGB by using the command:
\begin{equation}\label{ffmpeg_cfg}
\begin{aligned}
\rm ffmpeg \ 
&\rm \verb|-|{r} \ [FPS] \ \verb|-|{s} \ [W]*[H] \ \verb|-|{pix\_fmt} \ yuv420p \\
&\rm \verb|-|{i} \ [IN].yuv \ [OUT].png
\notag
\end{aligned}
\end{equation}
Here, \textit{FPS} is the frame rate, \textit{W} is width, \textit{H} is height, \textit{IN} is the name of input file and \textit{OUT} is the name of output file. As mentioned in~\cite{Agustsson_2020_CVPR}, it is not ideal to evaluate the standard codecs in RGB color space because the native format of test sets are YUV420. To reduce this effect, we treat the RGB video frames as the source data and convert them into YUV444 as the input of the standard codecs. The reconstructed videos are converted back into RGB for evaluation. This kind of operation is commonly used in recent works of learned image compression \cite{balle2018variational, minnen2018joint}.

\begin{figure*}[htb]
\centering
\subfloat[\label{fig:cfg-HM}]{
    \includegraphics[width=0.8\linewidth]{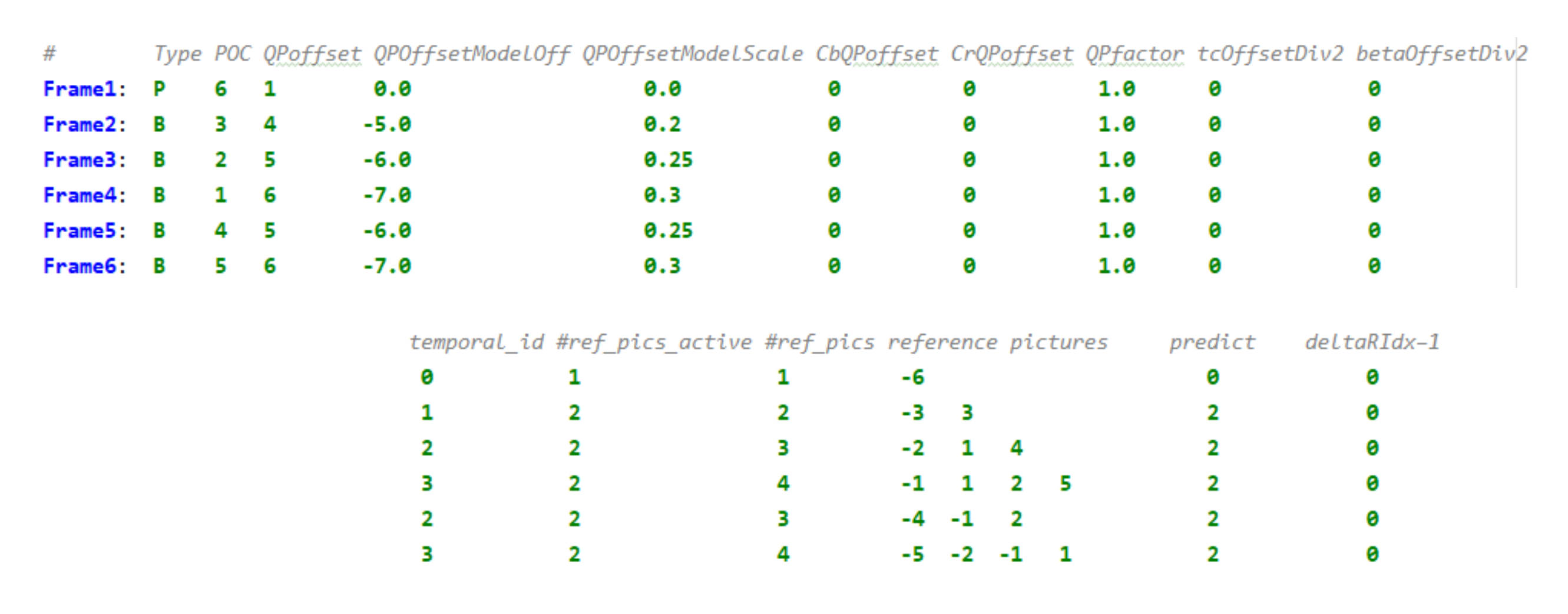}
}\\
\subfloat[\label{fig:cfg-VTM}]{
    \includegraphics[width=0.9\linewidth]{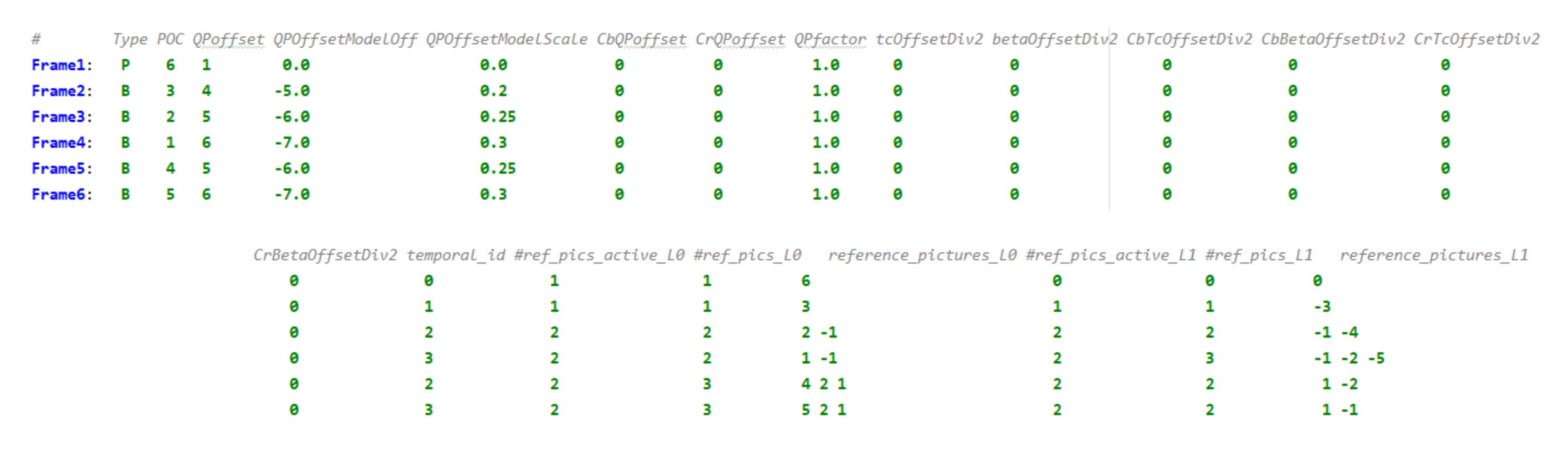}
}
\caption{The GOP structures of (a) HM and (b) VTM.}\label{fig:cfg}
\end{figure*}

\paragraph{HEVC reference software (HM)}
For lowdelay setting, we simply use the default ``encoder\_lowdelay\_P\_main.cfg'' configuration file of HM 16.21~\cite{HM}. For randomaccess setting, we change the gop structure of the default ``encoder\_randomaccess\_main.cfg'' configuration file, as shown in Fig.~\ref{fig:cfg}.
The following command is used to encode all HM videos:
\begin{equation}\label{HM_cfg}
\begin{aligned}
&\rm TAppEncoderStatic \ 
\rm \verb|-|{c} \ [CFG] \ \verb|-|{i} \ [IN].yuv \ \verb|-|{b} \ [OUT].bin\\ 
&\rm \verb|-|{o} \ [OUT].yuv \ \verb|-|{wdt} \ [W] \ \verb|-|{hgt} \ [H] \ \verb|-|{fr} \ [FPS] \ \verb|-|{f} \ [N] \\ 
&\rm \verb|-|{q} \ [QP] \ \verb|--|{IntraPeriod\verb|=|12} \ \verb|--|{Profile\verb|=|main\_444} \\
&\rm \verb|--|{InputChromaFormat\verb|=|444} \ \verb|--|{Level\verb|=|6.1} \\ 
&\rm \verb|--|{ConformanceWindowMode\verb|=|1}
\notag       
\end{aligned}
\end{equation}
Here, \textit{N} is the number of frames to be encoded for each sequence, which is set as 100 for the HEVC dataset and 600 for the UVG dataset.

\paragraph{VVC reference software (VTM)}
For randomaccess setting, we change the gop structure of the default \textit{encoder\_randomaccess\_main.cfg} configuration file of VTM 12.0~\cite{VTM}, as shown in Fig.~\ref{fig:cfg-VTM}. The following command is used to encode all VTM videos:

\begin{equation}\label{VTM_cfg}
\begin{aligned}
&\rm EncoderAppStatic \ 
\rm \verb|-|{c} \ [CFG] \ \verb|-|{i} \ [IN].yuv \ \verb|-|{b} \ [OUT].bin \\
&\rm \verb|-|{o} \ [OUT].yuv \ \verb|-|{wdt} \ [W] \ \verb|-|{hgt} \ [H] \ \verb|-|{fr} \ [FPS] \ \verb|-|{f} \ [N] \\
&\rm \verb|-|{q} \ [QP] \ \verb|--|{IntraPeriod\verb|=|12} \ \verb|-|{c} \ yuv444.cfg \\
&\rm \verb|--|{InputBitDepth\verb|=|8} \ \verb|--|{OutputBitDepth\verb|=|8} \\ 
&\rm \verb|--|{InputChromaFormat\verb|=|444} \ \verb|--|{Level\verb|=|6.1} \\ 
&\rm \verb|--|{ConformanceWindowMode\verb|=|1}
\notag       
\end{aligned}
\end{equation}
Here, \textit{N} is the number of frames to be encoded for each sequence, which is set as 100 for the HEVC dataset and 600 for the UVG dataset.

The default GOP structures of VLVC are almost the same as the structures used for HM and VTM, where the flow prediction module is turned on and the number of reference frames is set as 3.

\bibliography{reference}
\end{document}